\documentclass[onecolumn,superscriptaddress,aps,pre]{revtex4-1}
\usepackage{color}
\usepackage{amsmath}
\usepackage{amssymb}
\usepackage{graphicx}
\usepackage{subfigure}
\usepackage{ulem}

\usepackage{appendix}

\newcommand{\SI}{Supplementary Information}

\DeclareMathOperator*{\tr}{\text{tr}}

\newcommand{\mat}[1]{\boldsymbol{#1}}
\newcommand{\vect}[1]{\boldsymbol{#1}}
\DeclareMathOperator*{\ud}{\text{d}}

\newcommand{\quat}[1]{\boldsymbol{#1}}
\newcommand{\matquat}[1]{\tilde{\boldsymbol{#1}}}

\DeclareMathOperator*{\G}{\mathcal{G}}

\renewcommand{\emph}[1]{\textit{#1}}

\begin{document}


\title{The effect of population abundances on the stability of large random ecosystems}


\author{Theo Gibbs}
\affiliation{Dept.~of Ecology \& Evolution, University of Chicago. Chicago, IL 60637, USA}

\author{Jacopo Grilli}
\affiliation{Dept.~of Ecology \& Evolution, University of Chicago. Chicago, IL 60637, USA}

\author{Tim Rogers}
\affiliation{Centre for Networks and Collective Behaviour,
    Department of Mathematical Sciences, University of
    Bath. Claverton Down Bath BA2 7AY United Kingdom.}

\author{Stefano Allesina}
\affiliation{Dept.~of Ecology \& Evolution, University of Chicago. Chicago, IL 60637, USA}
\affiliation{ Computation Institute, University of Chicago.}
\affiliation{Northwestern Institute on Complex Systems (NICO),
  Northwestern University.}

\begin{abstract}
Random matrix theory successfully connects the structure of
interactions of large ecological communities to their ability to
respond to perturbations. One of the most debated
aspects of this approach is the missing role of population
abundances. Despite being one of the most studied patterns in ecology,
and one of the most empirically accessible quantities, population
abundances are always neglected in random matrix approaches and their
role in determining stability is still not understood.  Here, we
tackle this question by explicitly including population abundances in
a random matrix framework.  We obtain an analytical formula that
describes the spectrum of a large community matrix for arbitrary
feasible species abundance distributions. The emerging picture is
remarkably simple: while population abundances affect the rate to
return to equilibrium after a perturbation, the stability of large
ecosystems is uniquely determined by the interaction matrix.  We
confirm this result by showing that the likelihood of having a
feasible and unstable solution in the Lotka-Volterra system of
equations decreases exponentially with the number of species for
stable interaction matrices.
\end{abstract}

\maketitle

\section{Introduction}
Since the work of Lotka and Volterra, ecologists have attempted to
mathematize the interactions between populations to build predictive
models of population dynamics. This is a complex problem -- 
ecological communities are often composed of a large number of
species~\cite{May1988,May1988}, the equations describing
their interactions have been debated for
decades~\cite{Arditi2012}, and the estimation of parameters and
initial conditions is often unfeasible from an empirical standpoint.

To circumvent this problem, Robert May~\cite{May1972} introduced the
idea of modeling complex ecological communities using random
matrices. Consider the case in which the dynamics of the populations
can be described by a system of ordinary differential equations:

\begin{equation}
  \frac {dx_i(t)} {dt} = f_i(\vect{x}(t)),
  \label{eq:generic}
\end{equation}

\noindent where $\vect{x}(t)$ is a vector containing the populations
abundances at time $t$, and the function $f_i$ relates the abundance
of all populations to the growth of population $i$. In general, $f_i$
is a nonlinear equation with several parameters.

Suppose that the system admits a feasible equilibrium point, i.e., a
vector $\vect{x}^\ast$ such that $f_i(\vect{x}^\ast) = 0$ and
$x_i^\ast > 0$ for all $i$. If we start the system at this point, it
will remain there indefinitely. We can therefore ask whether the
system will go back to the equilibrium, or rather move away from it,
following a perturbation. This type of stability analysis can be
carried out by building the Jacobian matrix $J_{ij} = \partial
f_i(\vect{x}(t)) / \partial x_j$ and evaluating it at the equilibrium
point, yielding the so-called community matrix $\mat{M} = \left
. \mat{J} \right \rvert_{\vect{x}^\ast}$. If all the eigenvalues of
$\mat{M}$ have negative real part, then the equilibrium is locally
asymptotically stable, and the system will return to it after
sufficiently small perturbations; if any of the eigenvalues have a
positive real part, the system will move away from the equilibrium
when perturbed.

Clearly, to build $\mat{M}$ one would need to precisely know the
functions $f_i$, as well as their parameters, and solve for the
equilibrium (or equilibria) $\vect{x}^\ast$. May took a radically
different approach and analyzed the case in which $\mat{M}$ is a
random matrix with independent, identically distributed off-diagonal
elements, and constant diagonal elements~\cite{May1972}. For this
parameterization, he was able to show that the community matrices
describing sufficiently large and complex ecological communities are
always unstable. The random-matrix approach was recently extended and
refined to include different types of interaction between the
populations~\cite{Allesina2012,ReviewRMT}, as well as to study the
effect of more complex network structures, such as the hierarchical
organization of food-webs~\cite{Allesina2015a} and the modular pattern
often displayed by biological networks~\cite{Grilli2016}.

By modeling directly the matrix $\mat{M}$ as a random matrix, one does
not require a precise characterization of the functions $f_i$ and the
equilibrium $\vect{x}^\ast$. While mathematically convenient, this
approach does not explicitly take into account the abundance of the
populations---a type of data that is empirically much more accessible
than interaction coefficients or the elements of the community matrix.

The distribution of species abundances (SAD) has been shown to have
remarkably similar features across different species rich
communities~\cite{Bell2001a} with a skewed shape and few highly
abundant species. The log-series distribution~\cite{Fisher1943a},
discrete lognormal~\cite{Preston1948} and negative
binomial~\cite{Volkov2007} have all been proposed to describe
empirical SADs, and have been shown to emerge from either
neutral~\cite{Caswell1976,Hubbell2001a,Volkov2003,Azaele2016}
or niche mechanisms~\cite{MacArthur1957a,Vandermeer1966}.

The role of species abundances in structuring the community matrix
$\mat{M}$ can be easily seen by considering one of the simplest models
of population dynamics, the Generalized Lotka-Volterra (GLV) model:

\begin{equation}
\frac {dx_i(t)} {dt} = x_i(t) \left(r_i + \sum _{j} A_{ij} x_j(t) \right) \ ,
\label{eq:LV}
\end{equation}

\noindent where $r_i$ is the intrinsic growth rate of species $i$, and
$A_{ij}$ is the per-capita effect of species $j$ on the growth of
$i$. If a feasible equilibrium (i.e., one where all species have
positive abundance) exists, then it can be found solving the system of
equations

\begin{equation}
0 =  r_i + \sum _{j} A_{ij} x^\ast_j  \ ,
\end{equation}

\noindent yielding the community matrix $M_{ij} = A_{ij} x_i^\ast$,
which can be written in matrix form as

\begin{equation}
  \mat{M} = \mat{X} \mat{A} \ ,
\end{equation}

\noindent where $\mat{X}$ is a diagonal matrix with $X_{ii} =
x_i^\ast$ and zeros elsewhere. Even if the elements of $\mat{A}$ were
independent, identically distributed samples from a distribution, the
elements of $\mat{M}$ would not be---the matrix of abundances $\mat{X}$
couples all the coefficients in the same row, such that the
distribution of the elements in each row would in principle be
different.

One of the main goals of this work is to extend the random matrix
approach by considering a random matrix of abundances $\mat{X}$ and a
random matrix of interactions $\mat{A}$, and determining the stability
of $\mat{M}$ under these conditions. In this way, we address the
effect of species abundances on stability, thereby lifting one of the
main criticisms of the random matrix
approach~\cite{ReviewRMT,amnatjames,Jacquet2016}.

As we stated above when analyzing coexistence, we need population
abundances to be positive (\emph{feasible}). Stability cannot, at
least in principle, be disentangled from the constraint imposed by
feasibility on interactions~\cite{Roberts1974}. Diversity and
interaction properties have important consequences for the range of
parameters corresponding to feasible
solutions~\cite{Rohr2014,Stone2016,Grilli2017}. While the
interest in feasibility has grown considerably in recent years, the
relationship between feasibility and stability is still unclear. In
fact, most of the studies on feasibility assume strong conditions on
the interaction matrix (e.g., D-stability, diagonal stability) that
guarantee stability of any feasible
solution~\cite{Rohr2014,Grilli2017}. It is still unclear when these
assumptions are justified and how likely it is for large random
interaction matrices to meet these conditions.

In the second part of this work, we focus on the relationship between
feasibility and stability. In particular, we study the relationship
between the stability of $\mat{A}$ and that of $\mat{M}$ for the GLV
model. Our results show that, given a stable random matrix $\mat{A}$,
the probability that an arbitrary feasible equilibrium is unstable
decreases exponentially with diversity. This result strongly suggests
that, provided that the interaction matrix $\mat{A}$ is stable,
feasible solutions are almost surely stable. We therefore provide a
more robust justification to both May's original paper---by showing
that population abundances do not affect qualitatively stability---and
the more recent work on feasibility that assumes stability---by
predicting that this assumption is almost surely met for large random
systems.

\section{Constructing the community matrix with arbitrary population abundance}

We consider a system of $S$ interacting populations whose dynamics are
described by the GLV model in equation~\ref{eq:LV}, assume that a
feasible equilibrium $\vect{x}^\ast$ exists, and define $\mat{X}$ as
the diagonal matrix with diagonal entries $X_{ii} = x^\ast_i$. The
feasible fixed point $\vect{x}^\ast$ is locally asymptotically stable
if and only if all the eigenvalues of the community matrix $\mat{M} =
\mat{X} \mat{A}$, with components $M_{ij} = x^\ast_j A_{ij}$, have a
negative real part. Here, we model $\mat{A}$ as a random matrix and
$\vect{x}^\ast$ as a random vector with positive components, with the
goal of studying the spectrum (distribution of the eigenvalues) of the
community matrix $\mat{M} $ . From the GLV model, specifiying
a feasible fixed point $\vect{X}^\ast$ is the same as specifying a vector
of intrinsic growth rates $\vect{r}$ inside the feasibility domain~\cite{Rohr2014,Grilli2017}.

More specifically, we assume that the diagonal entries of the diagonal
matrix $\mat{X}$ are drawn from an arbitrary distribution with
positive support, mean $\mu_X$, and variance $\sigma_X^2$.  The
diagonal entries of $\mat{A}$ are drawn from an arbitrary distribution
with support in the negative axis, mean $\mu_d$, and variance
$\sigma_d^2$. Finally, each off-diagonal pair $(A_{ij},A_{ji})$ in
$\mat{A}$ is drawn independently from a bivariate distribution with
identical marginal means $\mu$, variances $\sigma^2$, and correlation
$\rho$. Unless otherwise specified, we focus on the case $\sigma_d =
0$, while we discuss in the \SI\ the effects of variability in
self-regulation.

\begin{figure}
\centering
  \includegraphics[width = 0.6\textwidth]{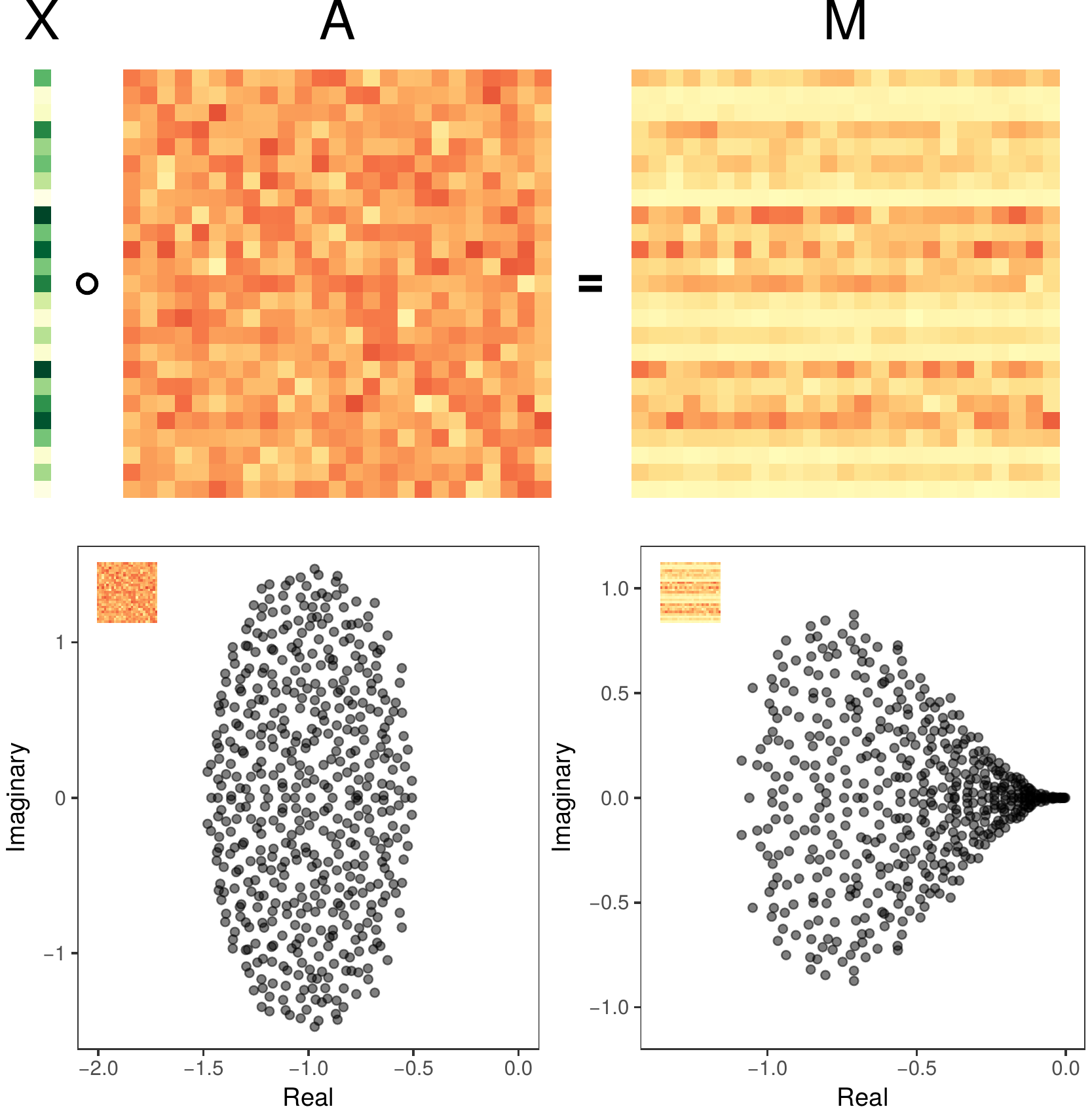}
	\caption{The top row shows the vector of abundances
          $\vect{x}^\ast$, the interaction matrix $\mat{A}$ and the
          community matrix $\mat{M} = \mat{X} \mat{A}$ (where
          $\mat{X}$ is a diagonal matrix with diagonal entries
          $\vect{x}^\ast$), with colors from red (negative) to green
          (positive). The bottom row shows the eigenvalues of
          $\mat{A}$ and $\mat{M}$, for $S = 500$. The diagonal entries
          of $\mat{X}$ are sampled from a uniform distribution on
          $[0,1]$, and matrix $\mat{A}$ is built sampling
          independently each pair $(A_{ij}, A_{ji})$ from a normal
          bivariate distribution with identical marginals defined by
          $\mu = 0$, $\sigma = 1 / \sqrt{S}$, and correlation $ \rho =
          -0.5$. The diagonal elements of $\mat{A}$ are fixed at -1.
          The main goal of this work is to characterize the spectrum
          of $\mat{M}$ given the properties of $\mat{A}$ and
          $\mat{X}$.}
	\label{fig:populationaffect}
\end{figure}

In the case of $\sigma_d = 0$ and in the limit of large $S$, the
spectrum of $\mat{A}$ is known and is independent of the choice of the
bivariate distribution (provided that mild conditions on the
finiteness of the moments are satisfied~\cite{nguyen2012elliptic}).
In particular, $\mat{A}$ has one eigenvalue equal to $-\mu_d + S
\mu$~\cite{ORourke2014}, while the others (the \emph{bulk} of
eigenvalues) are uniformly distributed in an ellipse in the complex
plane centered in $-\mu_d-\mu$ with horizontal axis $\sqrt{S} \sigma
(1+\rho)$ and vertical axis $\sqrt{S} \sigma
(1-\rho)$~\cite{nguyen2012elliptic,Allesina2012,ORourke2014}.
Figure~\ref{fig:populationaffect} shows an example of the spectrum of
$\mat{A}$.

Figure~\ref{fig:populationaffect} also shows an example of the
eigenvalues of the community matrix $\mat{M} = \mat{X} \mat{A}$ where
the diagonal entries of $\mat{X}$ are independent random variables
drawn from a uniform distribution. It is evident that the bulk of
eigenvalues of $\mat{M}$ does not follow the elliptic law.

\section{Disentangling the effect of the mean interaction strength}
\label{sec:disentangle}

When the mean $\mu$ of the off-diagonal elements of the interaction
matrix $\mat{A}$ does not equal zero, the spectra of $\mat{A}$ and
$\mat{M}$ are characterized by the presence of an outlier. The value
of this eigenvalue for the matrix $\mat{A}$ is known for the case
$\sigma_d = 0$, and in the limit of large $S$~\cite{ORourke2014}. It
can be obtained by decomposing the matrix $\mat{A}$ as a sum of three
matrices

\begin{equation}
\mat{A} = (\mu_d-\mu) \mat{I} + \mu \mat{1} + \mat{B} \ ,
\end{equation}

\noindent where $\mat{I}$ is the identity matrix, $\mat{1}$ is a
matrix of ones, and $\mat{B}$ is a random matrix with mean zero that
follows the elliptic law. It has been proved~\cite{ORourke2014} that the
spectrum of $\mat{A}$ is characterized by a bulk of eigenvalues,
determined by the spectrum of $(\mu_d-\mu) \mat{I} + \mat{B}$, and the
presence of an outlier, whose value is (approximately) given by the
largest eigenvalue of $(\mu_d-\mu) \mat{I} + \mu \mat{1}$, which has
value $\mu_d + (S-1) \mu$.

\begin{figure}
\centering
  \includegraphics[width = 0.8\textwidth]{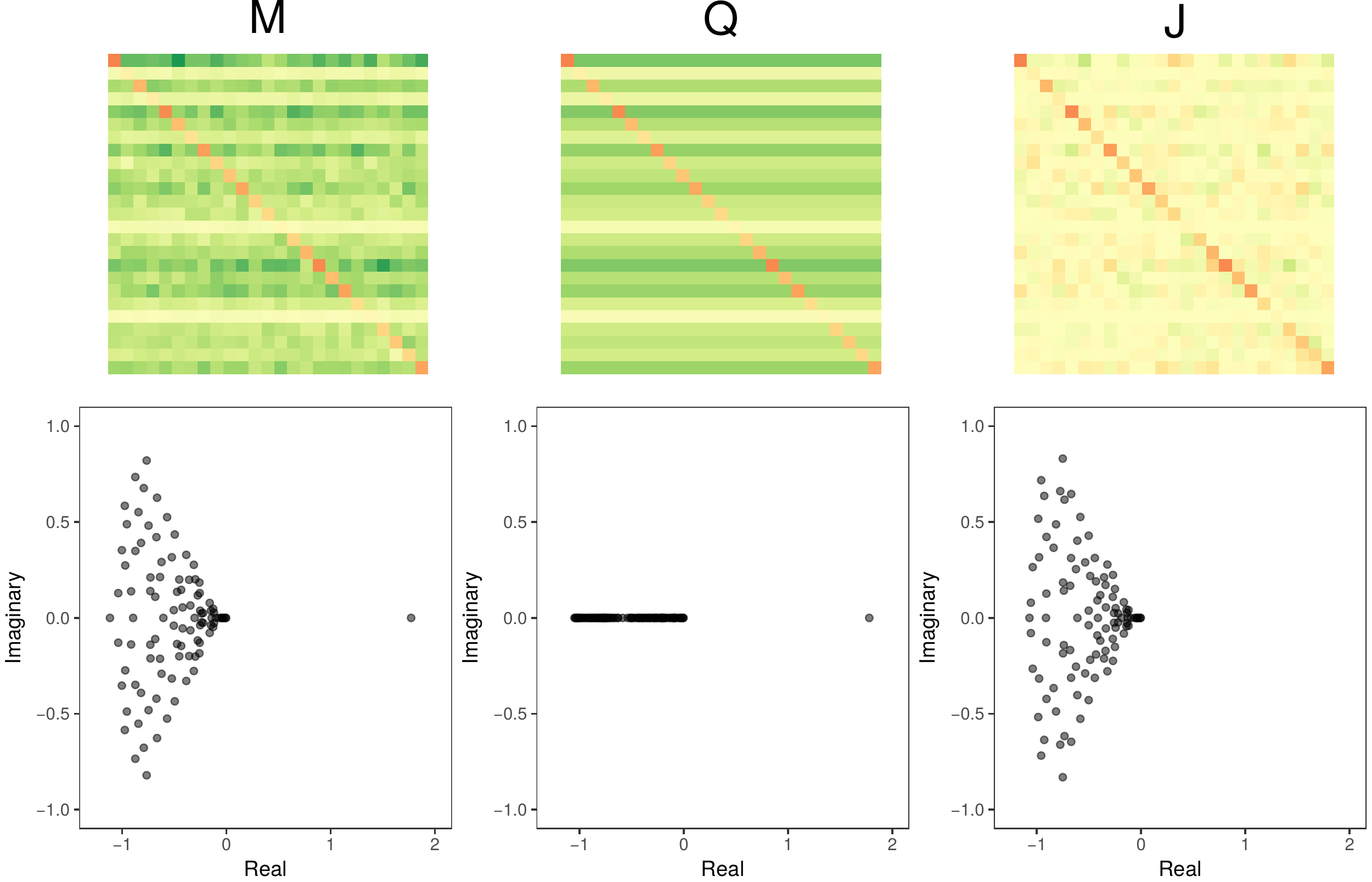}
	\caption{The top row shows the three matrices $\bf{M}$,
          $\bf{Q}$ and $\bf{J}$. The community matrix $\mat{M} =
          \mat{X} \mat{A}$, is obtained from the interaction matrix
          $\mat{A}$ that, without loss of generality, can be written
          as $\mat{A} = (\mu_d - \mu) \mat{I} + \mu \mat{1} + \mat{B}$,
          where $\mat{1}$ is a matrix of ones and $\mat{B}$ is a random
          matrix with diagonal elements fixed at zero whose coefficients
          have mean zero and variance $\sigma^2$. We define $\mat{Q} =
          \mat{X} ( (\mu_d - \mu)\mat{I} + \mu \mat{1}) $ and $\mat{J} = \mat{X}
          (\mu_d\mat{I} + \mat{B})$.  Equivalently, $\mat{Q}$ is the matrix
          with the same parameters as $\mat{M}$ except with $\sigma =
          0$, and $\mat{J}$ is obtained from the same parameters as
          $\mat{M}$ except with $\mu = 0$ for the off-diagonal terms.
          Remarkably, the eigenvalues of $\mat{M}$, $\mat{J}$ and
          $\mat{Q}$ are simply related: the bulk of eigenvalues of
          $\mat{J}$ and that of $\mat{M}$ are the same, while the
          outlier of $\mat{M}$ is the same as that of $\mat{Q}$. This
          decomposition allows us to obtain an analytical prediction
          for the outlier, and in the \SI\ we find the spectrum of
          $\mat{Q}$ analytically. In the figure, we set $S = 500$. The
          diagonal entries of $\mat{X}$ are sampled from a uniform
          distribution on $[0,1]$. $\mat{A}$ is sampled from a normal
          bivariate distribution with identical marginals $\mu =
          5 / S$, $\sigma = 5 / \sqrt{S}$ and correlation $\rho =
          -0.5$. }
	\label{fig:observation}
\end{figure}

Figure~\ref{fig:observation} shows that, if $\mu \neq 0$, the spectrum
of $\mat{M}$ is also characterized by the presence of a bulk and of an
outlying eigenvalue. By decomposing the matrix $\mat{M}$ as

\begin{equation}
\mat{M} = \mat{X} \left( (\mu_d-\mu) \mat{I} + \mu \mat{1} + \mat{B} \right) \ ,
\end{equation}

\noindent we show in the \SI\ that the bulk of the spectrum of
$\mat{M}$ is determined by the eigenvalues of the matrix $\mat{J} =
\mat{X} \left( (\mu_d-\mu) \mat{I} + \mat{B} \right)$ and the outlier
is given by largest eigenvalue of $\mat{Q} = \mat{X} \left(
(\mu_d-\mu) \mat{I} + \mu \mat{1} \right)$.
Figure~\ref{fig:observation} shows an example of this decomposition,
where it is evident that the bulks of eigenvalues of $\mat{M}$ and
$\mat{J}$ are the same, and the outliers of $\mat{M}$ and $\mat{Q}$
match.

The trace of $\mat{M}$ is given by

\begin{equation}
\tr\left(\mat{M}\right) = \lambda_{\textrm{out}} + (S-1) \langle \lambda
\rangle_{\textrm{bulk}} \ ,
\end{equation}

\noindent where $\lambda_{\textrm{out}}$ is the value of the outlier and
$\langle \lambda \rangle_{\textrm{bulk}}$ is the average eigenvalue in the
bulk.  Since the bulks of the eigenvalues of $\mat{M}$ and $\mat{J}$
are the same, we have that

\begin{equation}
\langle \lambda \rangle_{\textrm{bulk}}  =  \frac{1}{S} \tr\left(\mat{J}\right) = \mu_X \left(\mu_d - \mu \right)  \ .
\end{equation}

Using the fact that

\begin{equation}
\tr\left(\mat{M}\right) = S \mu_d \mu_X  \ ,
\end{equation}

\noindent we see that the outlier is equal to

\begin{equation}
\lambda_{\textrm{out}} = \mu_X \left( \mu_d + (S-1) \mu \right) \ .
\label{eq:outlier}
\end{equation}

\begin{figure}
\centering
  \includegraphics[width = 0.75\textwidth]{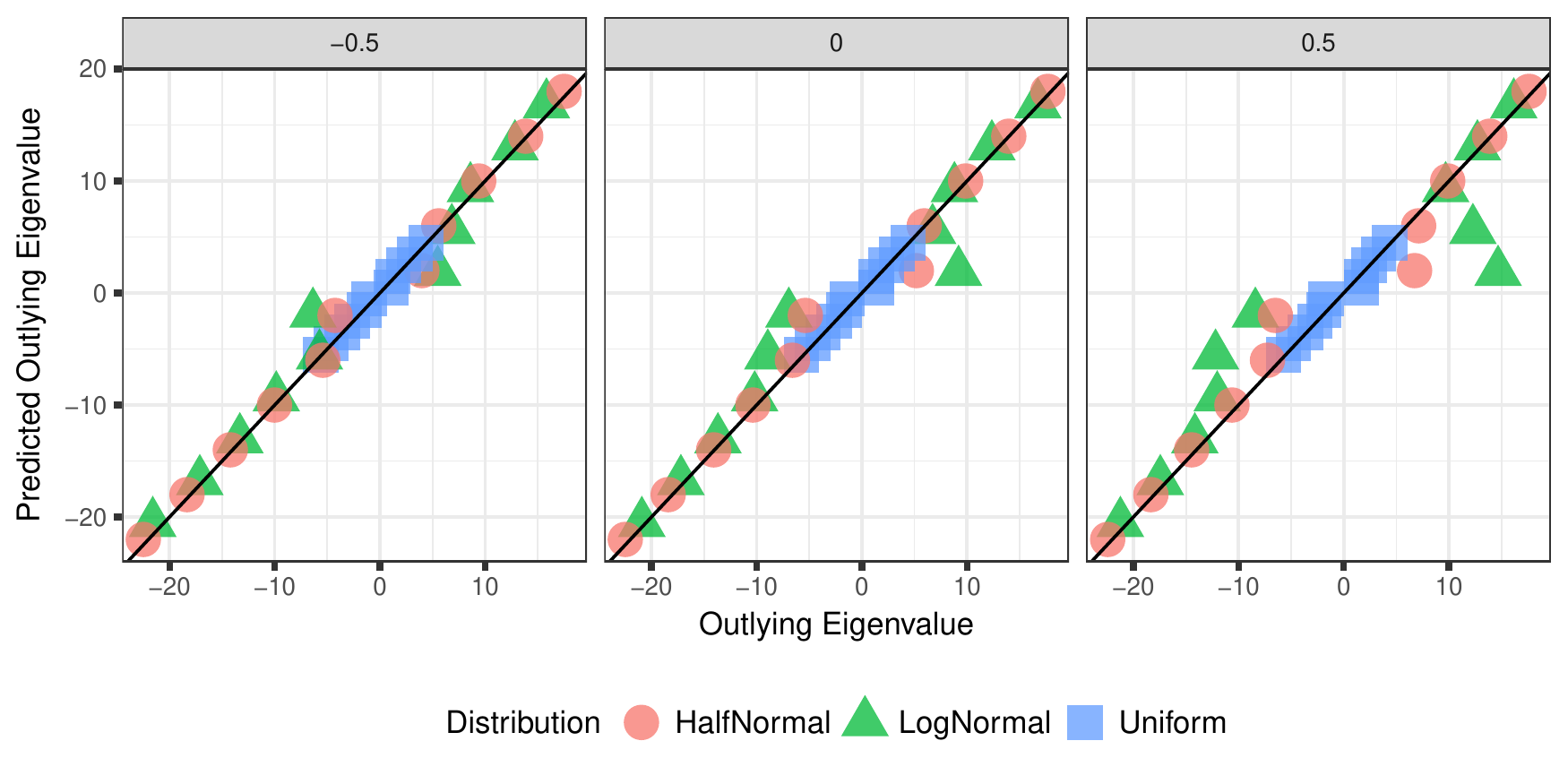}
	\caption{The three panels show that our analytical prediction
          (equation~\ref{fig:outeigen}) correctly matches the outlier
          of the spectrum, for different values of $\rho$ and
          population abundance distributions. The matrix $\mat{A}$ is
          built independently sampling the coefficients from a normal
          bivariate distribution with identical marginals defined by
          $\mu$, $\sigma$, and $\rho$. Here, we set $S = 1000$ and
          $\sigma = 1 / \sqrt{S}$, and vary $\mu$ between $-10$ and
          $10$ to test our prediction. We draw $\mat{X}$ from three
          different distributions with positive support: uniform (on
          [0,1]), log-normal (with mean log-mean $0.5$ and
          log-standard deviation $0.5$) and half-normal (shifted
          rightwards to have support $(1, \infty)$, and with parameter
          $ \theta = 1$).  We can observe deviations from our
          prediction when $\mu$ is small, especially when $\mat{X}$ is
          drawn from a log-normal distribution. This is because the
          eigenvalue corresponding to equation~\ref{eq:outlier}
          is now contained in the bulk.}
	\label{fig:outeigen}
\end{figure}

Figure~\ref{fig:outeigen} shows that this analytical prediction
closely matches the outlier of the spectrum of $\mat{M}$.

\section{Analytical solution in the case $\rho = 0$}

In section~\ref{sec:disentangle} we showed that the spectrum of
$\mat{M}$ is characterized by a bulk of eigenvalues and an outlier,
which is determined by the mean of interaction matrix $\mu$. In the
following, we focus on the bulk of eigenvalues, so we assume $\mu
= 0$.

Using the cavity method~\cite{Rogers2008,Rogers2009,Grilli2016}, we
derive in the \SI\ a system of equations for the spectral density of
the matrix $\mat{M}$.  These equations cannot be explicitly solved in
the most general case, but they take a particularly simple form in the
case where the correlation $\rho = 0$.  In this case, it is possible
to write an implicit equation for the support of the spectrum, which
takes the form

\begin{equation}
\int {\ud} x \ {\ud} s \ P_{XD}(x,s) \ \frac{ S x^2 \sigma^2} {| \lambda - s
  x |^2 } = 1 \ ,
\label{eq:uncpred}
 \end{equation}

\noindent where $P_{XD}(x,s)$ is the joint distribution of the
population abundances $x$, with mean $\mu_X$ and variance
$\sigma_X^2$, and the self-regulation terms (i.e., the diagonal
elements of the interaction matrix) with mean $\mu_d$ and variance
$\sigma_d^2$.  The complex solutions $\lambda$ of this equation define
the support of the spectrum in the complex plane.  In the \SI\ we
explicitly solve the case of constant self-regulation terms (i.e.,
$\sigma_d = 0$) and population abundances drawn from a uniform
distribution.

\begin{figure}
\centering
  \includegraphics[width = 0.75\textwidth]{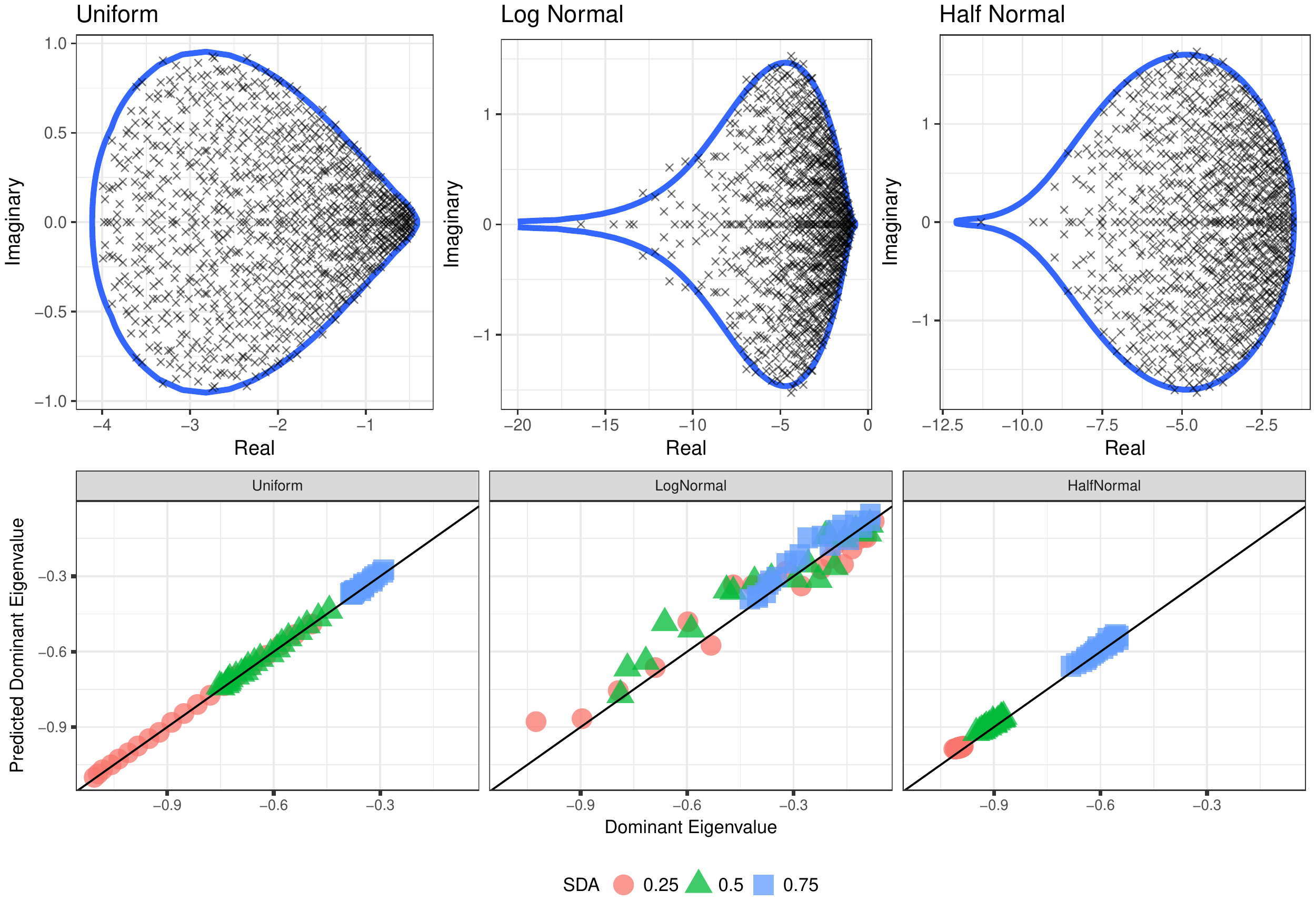}
	\caption{The top row shows that the analytical predictions for
          the support of the eigenvalue distribution obtained in
          equation~\ref{eq:uncpred_diag} (solid blue line) correctly
          predict the support of the spectrum of $\mat{M} = \mat{X}
          \mat{A}$.  In all the three plots, $\mat{A}$ is built using
          a bivariate normal distribution with identical marginals
          $\mu = 0$, $\sigma = 1 / \sqrt{S}$ and correlation $ \rho =
          0$. The diagonal entries of $\mat{A}$ are fixed at $-2$. We
          considered three different abundance distributions: uniform
          ($\mat{X}$ is sampled from a uniform distribution on [0.25,
            1.75]) lognormal ($\mat{X}$ is sampled from a log-normal
          distribution with log-mean $0.5$ and log-standard deviation
          $0.5$) and half-normal ($\bf{X}$ is sampled from a
          half-normal, shifted rightwards to have support $(1,
          \infty)$, and with parameter $ \theta = 1$). The bottom row
          shows the value of the rightmost eigenvalue of $\bf{M}$
          against the analytical prediction for the leading eigenvalue
          of matrices with the same abundance distributions used
          above, but varying their variances $\sigma_X^2$. Different
          colors correspond to different values of $\sigma$. Each
          point is an average over $20$ simulations.}
	\label{fig:uncorrelatedpredictions}
\end{figure}

When the self-regulation terms are constant, equation~\ref{eq:uncpred}
reduces to

\begin{equation}
\int {\ud}x \ P_{X}(x) \ \frac{S x^2 \sigma^2} {| \lambda - \mu_d x |^2 }  = 1 \ ,
\label{eq:uncpred_diag}
 \end{equation}

\noindent where $P_{X}(x)$ is the species abundance distribution.
Figure~\ref{fig:uncorrelatedpredictions} compares the analytical
prediction with the bulk of eigenvalues of $\mat{M}$ for different
distributions of $\mat{X}$, showing that the solutions
of equation~\ref{eq:uncpred_diag} closely match the support of the spectrum of
$\mat{M}$.

Equation~\ref{eq:uncpred_diag} also predicts that if $\mat{A}$ is
stable, then $\mat{M}$ is stable. In fact, equation~\ref{eq:uncpred_diag}
predicts  that the matrix $\mat{A}$ is
stable iff $\mu_d + S \sigma^2 < 0$. If this condition is met, it is
simple to observe that

\begin{equation}
\frac{S x^2 \sigma^2} {| \lambda - \mu_d x |^2 }  < 1
 \end{equation}

\noindent for any complex $\lambda$ with positive real part and any
positive real $x$. When this inequality is used in
equation~\ref{eq:uncpred_diag} one obtains that the points on the
boundary of the support, and therefore all the eigenvalues, always
have negative real part.

\section{The stability of large community matrices does not depend on population abundance}
\label{sec:dstab}

In the previous section, we derived the spectrum in the case $\rho =
0$, finding that if the interaction matrix $\mat{A}$ is stable, then
$\mat{M}$ is stable. The goal of this section is to study more deeply
the relationship between the stability of $\mat{A}$ and that of
$\mat{M}$. More specifically, given a stable random matrix $\mat{A}$,
we ask what is the probability of finding a positive diagonal matrix
$\mat{X}$, such that $\mat{M} = \mat{X} \mat{A}$ is stable.

A matrix $\mat{A}$ is \emph{D-stable} if, for any positive diagonal
matrix $\mat{X}$, $\mat{X} \mat{A}$ is
stable~\cite{Kaszkurewicz2000}. An explicit condition for D-stability
that does not require checking all the possible choices of $\mat{X}$
is not known in dimension larger than
four~\cite{Redheffer1985}. Therefore, it is not known, in general,
under which values of $\mu$, $\sigma$, $\rho$ and $\mu_d$ random
matrices are expected to be \emph{D-stable}.

A stronger condition for stability is \emph{diagonal stability}. A
matrix $\mat{A}$ is diagonally stable if there exists a positive
diagonal matrix $\mat{X}$ such that $\mat{X}\mat{A}+\mat{A}^t\mat{X}$
is stable.  Interestingly, diagonal stability implies
D-stability~\cite{Kaszkurewicz2000}. As for D-stability, a simple necessary
and sufficient test for diagonal stability is not known.  On the other hand, it is
simple to observe that the stability of $(\mat{A}+\mat{A}^t)/2$ is a
sufficient condition for diagonal stability (corresponding to choosing
a constant diagonal matrix $\mat{X}$), and therefore also implies
D-stability.

All the eigenvalues of $(\mat{A}+\mat{A}^t)/2$ are real and, if
$\mat{A}$ is a symmetric random matrix of independently distributed
entries with bounded higher moments, the bulk of eigenvalues of
$(\mat{A}+\mat{A}^t)/2$ follows Wigner's semicircle
distribution~\cite{Wigner1958,Tang2014}

\begin{equation}
\varrho_{\frac{\mat{A}+\mat{A}^t}{2}}(\lambda) = \frac{\sqrt{ 2 S \sigma^2 (1+\rho) - \left( \lambda - (\mu_d - \mu) \right)^2 }}{ \pi S \sigma^2 (1+\rho)} \ ,
\end{equation}

\noindent with one outlying eigenvalue equal to $\mu_d + (S-1) \mu$.

For positive mean $\mu$, if $\mu > (1+\rho) \sigma / \sqrt{S}$, the
rightmost eigenvalue is the outlier. In this case, the rightmost
eigenvalue of $\mat{A}$ and of $(\mat{A}+\mat{A}^t)/2$ are the
same. Therefore, for non-negative $\mu$, stable random matrices are almost
surely diagonally stable. Since diagonal stability implies
D-stability, if $\mat{A}$ is stable, then $\mat{M}=\mat{X}\mat{A}$ is
stable. This argument is in agreement with our formula for the outlier
of $\mat{M}$ in the case of non-vanishing mean $\mu$, obtained in
equation~\ref{eq:outlier}. For positive mean $\mu$, the rightmost
eigenvalue of $\mat{M}$ is equal to $\mu_X \lambda_{\mat{A}}$, where
$\lambda_{\mat{A}}$ is the rightmost eigenvalue of $\mat{A}$ and
$\mu_X$ is positive by definition. The sign of the rightmost
eigenvalue of $\mat{M}$ is therefore the same as that of the rightmost
eigenvalue of $\mat{A}$.

Since a negative $\mu$ only produces a equal shift in the rightmost
eigenvalue of $\mat{A}$, $(\mat{A}+\mat{A}^t)/2$ and $\mat{M}$, we can
restrict our analysis to the case $\mu = 0$. For vanishing mean, the
rightmost eigenvalue of $(\mat{A}+\mat{A}^t)/2$ is equal
to~\cite{Tang2014}

\begin{equation}
\lambda_{\frac{\mat{A}+\mat{A}^t}{2}} = \mu_d + \sqrt{2 S \sigma^2 (1+\rho) } \ ,
\label{eq:lowerdiag}
 \end{equation}

\noindent which should be compared with the rightmost eigenvalue of
$\mat{A}$

\begin{equation}
\lambda_{\mat{A}} = \mu_d + \sqrt{S \sigma^2} (1+\rho)  \ .
\label{eq:stab}
 \end{equation}

As shown in~\cite{Tang2014,Grilli2017},
$\lambda_{\frac{\mat{A}+\mat{A}^t}{2}} \geq \lambda_{\mat{A}}$ and
they are equal in the case $\rho = 1$.  Equation~\ref{eq:lowerdiag}
imposes a sufficient condition on diagonal stability: if

\begin{equation}
\mu_d + \sqrt{2 S \sigma^2 (1+\rho) } < 0 \ ,
\label{eq:diagstabsuff}
 \end{equation}

\noindent $\mat{A}$ is diagonally stable and, for any choice of
positive diagonal matrix $\mat{X}$, $\mat{M}=\mat{X}\mat{A}$ is
stable.  The non-trivial regime therefore corresponds to the values of
parameters where $\mu_d + \sqrt{2 S \sigma^2 (1+\rho) } > 0$ and
$\mu_d + \sqrt{S \sigma^2} (1+\rho) < 0$~\cite{Grilli2017}.

\begin{figure}
  \includegraphics[width = 0.75\textwidth]{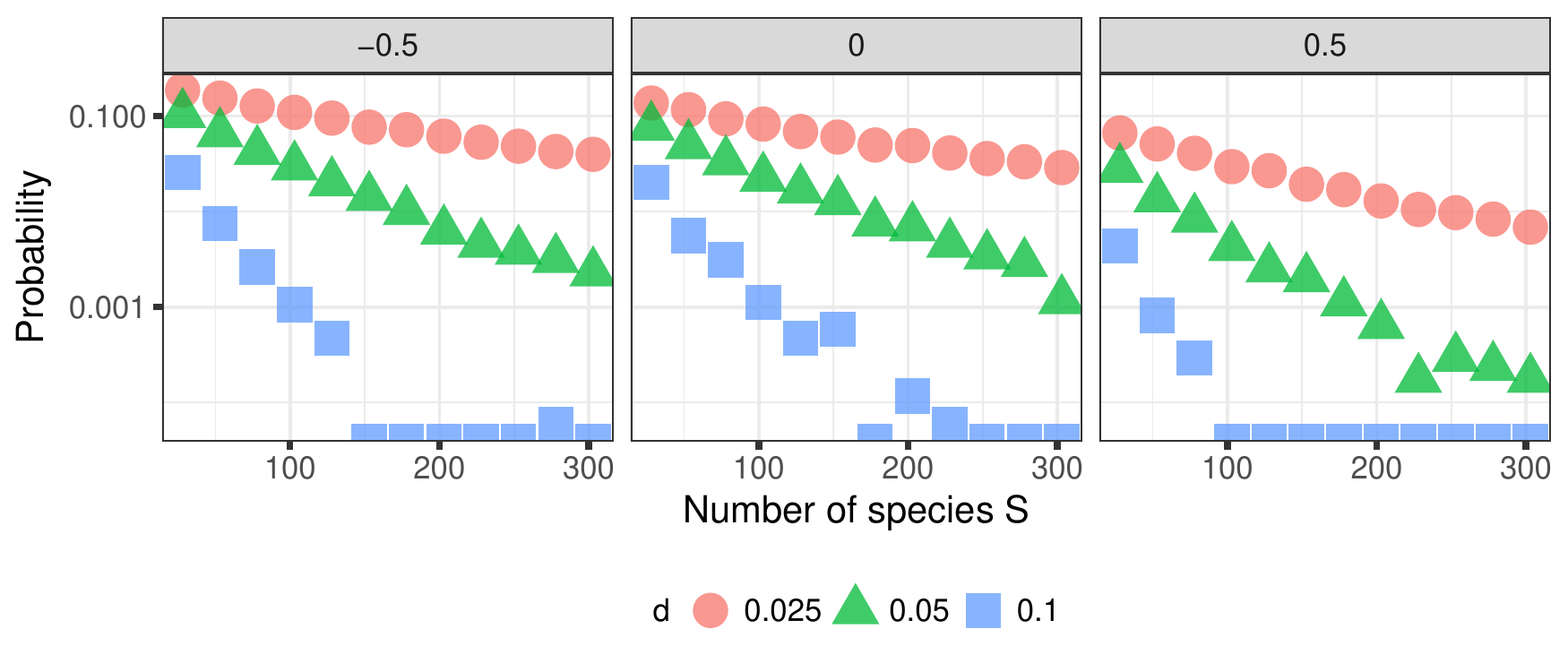}
	\caption{ We computed the probability that a matrix $\mat{M} =
          \mat{X} \mat{A}$ is unstable (i.e., that the leading
          eigenvalue has positive real part), given that $\mat{A}$ is
          a stable random matrix with rightmost eigenvalue equal to
          $\lambda_{max} = -d$. This probability decreases
          exponentially with $S$ for different values of $\rho$
          ($-0.5$ in the left panel, $0$ in the center and $0.5$ in
          the right) and $\lambda_{max}$ (different colors).  For a
          given number of species $S$, we construct the random matrix
          $\mat{A}$ sampling its entries from a bivariate normal with
          identical marginals $\mu = 0$, $\sigma = 1 / \sqrt{S}$ and
          given $\rho$.  The diagonal elements of $\mat{A}$ are all
          equal and their value is determined in order to have
          dominant eigenvalue equal to $\lambda_{max}$. The diagonal
          entries of $\mat{X}$ were sampled from a uniform
          distribution on $[0,1]$.  For each value of the parameters
          $\rho$, $\lambda_{max}$ and $S$, we constructed $XXX$
          matrices $\mat{A}$ and $\mat{X}$ and computed the fraction
          of matrices $\mat{M} = \mat{X}\mat{A}$ with positive
          dominant eigenvalue.  }
	\label{fig:unweightedprobabilities}
\end{figure}

Since an explicit condition for D-stability does not exist, we
computed the probability that, given a stable random matrix $\mat{A}$,
a positive diagonal matrix $\mat{X}$ would make $\mat{M} = \mat{X}
\mat{A}$ unstable.  Note that, since any matrix has a non-null
probability of being generated when entries are sampled from a
bivariate distribution with infinite support, this probability is
always non-zero. The relevant question in this context is therefore
how this probability depends on the number of species $S$.
Figure~\ref{fig:unweightedprobabilities} shows that the probability of
finding a $\mat{X}$ with a destabilizing effect decreases
exponentially with the number of species $S$, with a rate that depends
on the rightmost eigenvalue $\lambda_{\mat{A}}$ and the correlation
$\rho$. This implies that, for large values of $S$, $\mat{M}$ is
almost surely stable if $\mat{A}$ is stable.

\section{Fixed points are almost surely stable in large random Lotka-Volterra equations.}

If we consider the Lotka-Volterra equations
(equation~\ref{eq:LV}), and we set the values of the intrinsic growth
rates $\vect{r}$, the fixed point has components

\begin{equation}
x_i^\ast = \sum_j A^{-1}_{ij} r_j \ .
\label{eq:fixedpoint}
\end{equation}

Let us also assume that all these components are positive (i.e.,
$\vect{r}$ is inside the feasibility domain).  In
section~\ref{sec:dstab} we showed that the matrix obtained by
multiplying a stable random matrix $\mat{A}$ and a random positive
diagonal matrix $\mat{X}$ is more and more likely to be stable as $S$
increases.  It is evident (from equation~\ref{eq:fixedpoint}) that the
components of $\vect{x}^\ast$ are not independent of the entries of
the matrix $\mat{A}$.  The presence of this correlation implies that,
at least in principle, choosing a random vector $\vect{r}$ inside the
feasibility domain to define $\mat{X}$
could produce different results from
sampling independent entries from a specified species abundance
distribution.

In this section we repeat the simulations detailed in
section~\ref{sec:dstab}, but instead of considering a random fixed
point $\vect{x}^\ast$, we find the $\vect{x}^\ast$ determined by
a random intrinsic growth rate vector
$\vect{r}$ sampled uniformly from the feasibility domain.
The most intuitive method for this simulation would consist of taking
a random matrix $\mat{A}$, choosing a value $\vect{r}$ at random on
the unit sphere, checking if it corresponds to a feasible fixed-point
using equation~\ref{eq:fixedpoint}, and finally computing the
eigenvalue of $\mat{M} = \mat{X}\mat{A}$. However, as the number of
species $S$ increases this method becomes practically unfeasible. In
fact, the fraction of intrinsic growth rate vectors $\vect{r}$
corresponding to a feasible solution decreases exponentially with
$S$~\cite{Grilli2017}. If this intuitive method was employed, most of
the simulation time would be spent trying to find vectors $\vect{r}$
inside the feasibility domain.

On the other hand, since the relation between $\vect{r}$ and
$\vect{x}^\ast$ (via equation~\ref{eq:fixedpoint}) is bijective, we
can easily construct all the vectors $\vect{r}$ inside the feasibility
domain by considering all the possible feasible solution
$\vect{x}^\ast$. In section~\ref{sec:dstab} we specified a
distribution on the $\vect{x}^\ast$. This distribution translates to a
non-trivial distribution on the $\vect{r}$ (that can be obtained from
equation~\ref{eq:fixedpoint}). In this section, we instead assume a
distribution on the $\vect{r}$ and derive a corresponding distribution
for the $\vect{x}^\ast$. For instance, if we assume that the vectors
$\vect{r}$ are uniformly distributed on the unit sphere, the
distribution of the $\vect{x}^\ast$ reads~\cite{Grilli2017}

\begin{equation}
P(\vect{x}^\ast | \mat{A} ) \propto | \det \mat{A} | \frac{ \delta
  \left( \|\vect{x}^\ast\|^2 - 1 \right) }{ \| \mat{A} \vect{x}^\ast
  \|^S } \ .
\label{eq:probx}
 \end{equation}

Sampling vectors $\vect{x}^\ast$ according to this distribution is
equivalent to sampling vectors $\vect{r}$ uniformly from the
feasibility domain. It is important to observe that when
$\vect{x}^\ast$ is drawn according to this distribution, its entries
are not independent and their densities depend on $\mat{A}$.

\begin{figure}
  \includegraphics[width = 0.75\textwidth] {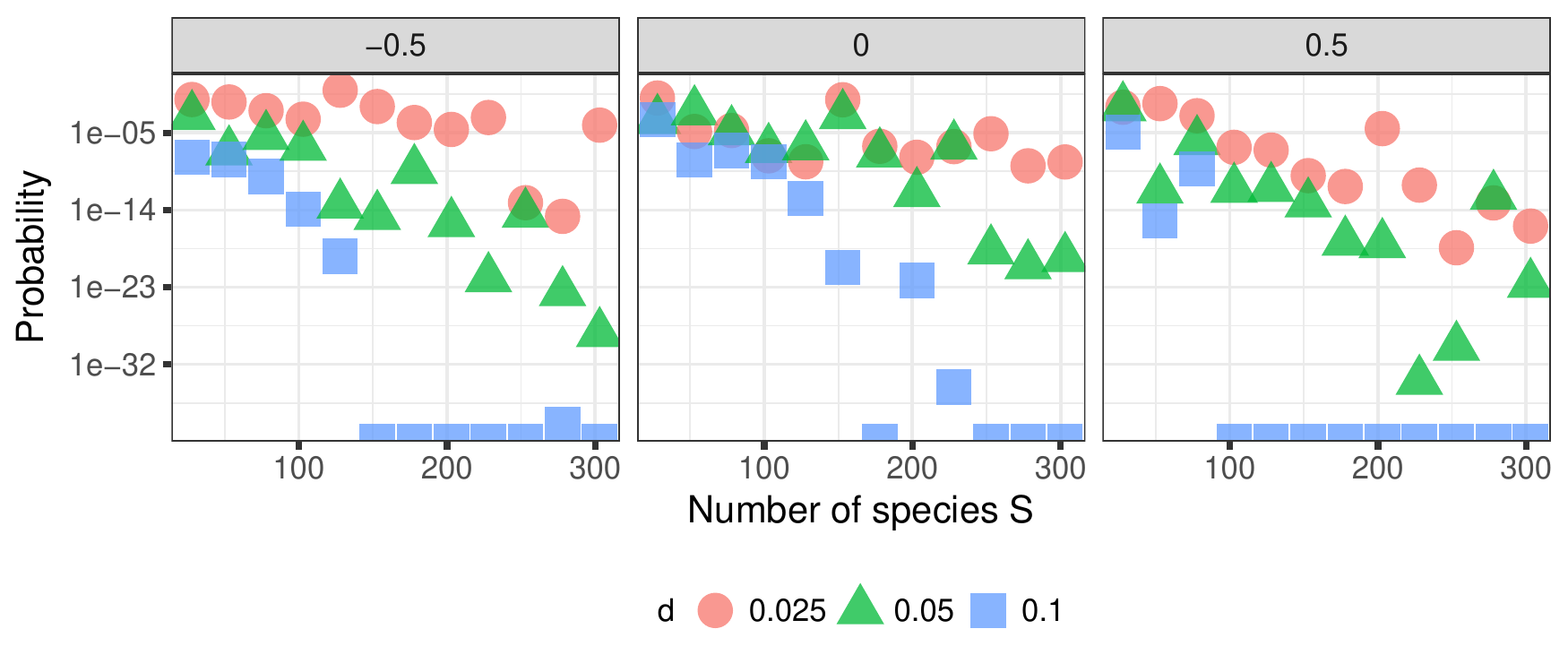}
	\caption{ These panels plot the same quantity as of
          fig~\ref{fig:unweightedprobabilities}. Instead of sampling
          $\mat{X}$ from a uniform distribution, we used the
          distribution of eq.~\ref{eq:probx}, which guarantee an
          unbiased sampling of the intrinsic growth rates of a
          Lotka-Volterra systems.  This sampling method is in fact
          equivalent to sampling a random interaction matrix $\mat{A}$
          and an intrinsic growth rate vector $\vect{r}$ inside the
          feasibility domain and checking the stability of the
          corresponding feasible fixed point. The exponential decay
          with increasing $S$ strongly suggests that the set of
          feasible unstable fixed points has measure zero for large,
          randomly interacting Lotka-Volterra systems.  }
	\label{fig:weightedprobabilitiess}
\end{figure}

Figure~\ref{fig:weightedprobabilitiess} shows the stability of
$\mat{M} = \mat{X} \mat{A}$ when the diagonal entries of $\mat{X}$ are
sampled from the probability distribution defined in
equation~\ref{eq:probx}.  Despite the presence of a correlation
between the entries of $\mat{X}$ and $\mat{A}$, the result obtained in
section~\ref{sec:dstab} is confirmed: the probability of observing a
stable $\mat{A}$ but an unstable $\mat{M}$ decreases exponentially
with $S$. If the interaction matrix $\mat{A}$ is stable, in the limit
of large $S$, the set of intrinsic growth rates corresponding to
feasible unstable solutions has measure zero.

\section{Discussion}

We explored the effect of population abundances on the stability of
random interacting ecosystems. We derived an expression for the
spectral density of a community matrix that explicitly includes the
species abundance distribution. While the effect on the eigenvalues is
highly heterogeneous and strongly depends on the specific choice of
the abundance distribution, a remarkably simple message emerges for
large randomly interacting ecosystems: the community matrix is stable
if and only if the interaction matrix is stable. In other words, the
abundances of species seem to not affect the sign of the eigenvalues.
We further explored this intriguing result by explicitly estimating
the probability of choosing a species abundance distribution leading
to instability. While for finite systems this probability is always
positive, it decreases exponentially with the number of species,
confirming what was found studying the spectrum of the community
matrix analytically.

Our results strongly suggest that large random matrices are D-stable
\emph{almost surely}: the set of destabilizing positive diagonal
matrices has measure zero. This fact has important consequences on
Lotka-Volterra systems of equations, implying that feasible
unstable fixed-points are very unlikely. This result allows to
disentangle the problem of feasibility (how often are fixed points
feasible?)  from the problem of stability (how often are fixed points
stable?), justifying a-posteriori what assumed in many studies on
feasibility~\cite{Rohr2014,Grilli2017} and expanding the validity of
their results.

The generalized Lotka-Volterra equations display a rich dynamical
behavior, leading to limit cycles when two or more species are
considered and chaos with three or more
species~\cite{Smale1976,Takeuchi1996}. Both limit cycles
and chaos require the existence of an unstable fixed point in the
interior of the feasibility
domain~\cite{Hofbauer1998}. Since the chance of observing
a feasible unstable fixed point decays rapidly when the number of
species increases, our results suggest that chaos and limit cycles are
extremely rare in large random Lotka-Volterra systems.

A stronger notion than D-stability is diagonal stability. While for
Lotka-Volterra systems, the former implies local asymptotic stability
of any feasible solution, the latter implies global stability.  We
showed that large random stable matrices are always D-stable. Under
which conditions they are also diagonally stable is an important open
problem. A sufficient condition for diagonal stability is negative
definiteness~\cite{Grilli2017}.  In the context of random matrices,
negative definiteness is equivalent to the condition expressed in
equation~\ref{eq:diagstabsuff}.  The condition for negative
definiteness should be compared to the condition for stability (see
equation~\ref{eq:stab}).  For large random matrices, two extreme
scenarios are possible: negative definiteness is almost surely a
necessary condition for diagonal stability, or stable random matrices
are almost surely diagonally stable. It
is also possible that the condition for diagonal stability is less
trivial, corresponding to values of parameters between the conditions
imposed by equations~\ref{eq:diagstabsuff} and \ref{eq:stab}.  Even
more complicated, it is also possible that a sharp condition for
diagonal stability does not exist for random matrices and, in the
limit of large $S$, stable and non negative definite random matrices
have a non-vanishing probability of being (or not being) diagonally
stable.

Our results shed light on one of the most controversial aspects of the
classic result of May~\cite{May1972} and its extensions.  Many
authors~\cite{Roberts1974,Pimm1979,King1983,ReviewRMT,amnatjames,Jacquet2016}
have argued that the unrealistic assumption of constant population
abundances was a key choice in May's paper, suggesting that more
realistic abundance distribution would have produced drastically
different results.  We showed that the conditions obtained
in the original paper and in its
extension~\cite{May1972,Allesina2012} are in fact valid for any
species abundance distribution.  In other words, the stability of
fixed points (i.e., the stability of the community matrix) is
determined only by the stability of the interaction matrix.

\begin{acknowledgments}
We thank A. Maritan, S. Tang and G. Barab\'as for comments and discussions.
T.G. and S.A. were supported by NSF grant
DEB-1148867. J.G. was supported by the Human
Frontier Science Program.
\end{acknowledgments}

\bibliographystyle{nature} \bibliography{abdrndmat}

\clearpage
\appendix

\setcounter{table}{0}
\renewcommand{\thetable}{A\arabic{table}}%
\setcounter{figure}{0}
\renewcommand{\thefigure}{A\arabic{figure}}%

\section{Notation and goals}

We aim to study the spectral density of a matrix $\mat{M}$ of the form $\mat{M = XA}$, where $\mat{X}$ is a positive diagonal matrix and $\mat{A}$ a random matrix with arbitrary distribution.
The diagonal entries of $\mat{X}$ are drawn from an arbitrary distribution with positive support, mean $\mu_X$ and variance $\sigma_X^2$.
The diagonal entries of $\mat{A}$ are drawn from an arbitrary distribution with negative support, mean $\mu_d$ and variance $\sigma_d^2$.
Each pair of off-diagonal entries $(A_{ij}, A_{ji})$ is drawn from a bivariate distribution with identical marginal means $\mu$, variances $\sigma^2$ and correlation $\rho$. 

Let $ \mat  B $ be an $ S \times S $ random matrix with complex eigenvalues $ \lambda_i$ for $i = 1, ... , S$. Its spectral density is defined as 
\begin{equation}
 \varrho(x,y) = \frac{1} {S} \sum_{i=1}^S\delta(x - \Re(\lambda_i))\delta(y - \Im(\lambda_i)) 
\end{equation}
which, in the limit of large $S$, converges to 
\begin{equation}
\varrho(x,y) = \mathbb{E} [\delta(x - \Re(\lambda_i))\delta(y - \Im(\lambda_i)) ] \ ,
\end{equation}
where $\mathbb{E}[\cdot]$ stands for the expectation over matrices in the ensemble.

We introduce the resolvent~\cite{Rogers2010}
\begin{equation}
\G(\quat{q}; \mat{B}) = \frac{1} {S} \sum_{i=1}^S (\lambda_i - \quat{q})^{-1} = \frac{1}{S} \tr\left( \mat{B} - \quat{q} \mat{I} \right)^{-1} \ .
\label{eq:resolvent_finite}
\end{equation}
The variable $\quat{q} = \lambda + \epsilon j$ is a quaternion (see section~\ref{sec:quaternions} for definitions and notation) and the resolvent is a function $\G : \mathbb{H} \to \mathbb{H}$.

The resolvent and the spectral density are related by the following formulas~\cite{Rogers2010}:
\begin{equation}
\G(\quat{q}; \mat{B}) = \int dx \ dy \ \varrho(x,y) (x+iy - \quat{q})^{-1} 
\label{resolvent+spectrum}
\end{equation}
and
\begin{equation}
\varrho\left(x,y \right) =  - \frac {1} {\pi} \lim_{\epsilon \to 0^+} \Re \left( \frac{\partial} {\partial \bar{ \lambda} }  \G(\lambda + \epsilon j; \mat{B})  \right) \biggl|_{\lambda = x+iy} \ ,
\label{spectrum+resolvent}
\end{equation}
where $\frac{\partial} {\partial \bar{ \lambda} }$ is the Wirtinger derivative
\begin{equation}
\frac{\partial }{\partial \bar{\lambda}} := \frac{1}{2} \left( \frac{\partial }{\partial x} + i \frac{\partial }{\partial y} \right) \ .
\end{equation}

\section{Quaternions}
\label{sec:quaternions}

When constructing the complex numbers from the real numbers, one defines a variable $i$ to be a root of the equation $x^2 + 1 = 0$.
The algebraic structure of $\mathbb{C}$ descends from the equation $i^2 = -1$ and the algebraic structure of $\mathbb{R}$.
Similarly, the algebra of quaternions $\mathbb{H}$ can be defined by introducing the symbols $i$, $j$ and $k$ and the relations 
\begin{equation}
i^2 = j^2 = k^2 = ijk = -1 \ .
\end{equation}
From these equations all the multiplication rules can be obtained. In particular, it follows that multiplication in $\mathbb{H}$ is not commutative
(e.g. $ij = -ji$).

A quaternion $\quat{q}$ can be written as 
\begin{equation}
\textbf{q} = a + bi + cj + dk \ ,
\end{equation}
where $a,b,c,d \in \mathbb{R}$.
Equivalently, by introducing the two complex numbers $z = a + bi$ and $w = c + di$ and using $k = ij$, one can write
\begin{equation}
\textbf{q} = z + w j \ .
\end{equation}
Another equivalent way to represent quaternions is to write them in matrix form
\begin{equation}
\textbf{q} = \left( \begin{matrix} z & w \\ \bar{w} & \bar{z} \\ \end{matrix} \right) \ .
\end{equation}
It can be shown that, when written in this form, the multiplication rules of quaternions match the rules of matrix multiplication. In particular,
one has that
\begin{equation}
(z + w j)( u + v j ) = ( z u - w \bar{v} ) + (z v + w \bar{u} ) j \ .
\end{equation}
We also introduce the operation
\begin{equation}
(z + w j) \circ ( u + v j ) =  z u - w v j \ ,
\end{equation}
which, in matrix notation, corresponds to element-by-element multiplication.

The conjugate of a quaternion $\textbf{q} = z + wj$ is defined as $\bar{\quat{q}} = \bar{z} - wj$.
From this definition, one obtains the norm of a quaternion
\begin{equation}
| \quat{q} |^2 \equiv \quat{q} \bar{\quat{q}} = | z|^2 + |w|^2   \ ,
\end{equation}
and the inverse 
\begin{equation}
 \quat{q}^{-1} \equiv \bar{\quat{q}} \frac{1}{|\quat{q}|^2}   \ .
\end{equation}
Moreover, the real part of a quaternion is defined as
\begin{equation}
\Re(\quat{q}) \equiv \bar{\quat{q}} + \quat{q} = \Re(z) = a   \ .
\end{equation}

\section{Bulk and outliers of the spectrum of $\mat{M}$}
\label{sec:mudiff}

In this section, we want to show that the mean of $\mat{A}$ does not affect the bulk of eigenvalues of $\mat{M}$.
We decompose the matrix $\bf{M}$ as 
\begin{equation}
 \mat{M} = \mat{X} (\mat{D} - \mu \mat{I} + \mat{B} + \mu \mat{1} ) 
\label{eq:matrix_decomposed}
\end{equation}
 where $\mat{I}$ is the identity matrix, $\mat{1}$ is a matrix of ones and $\mat D $ is the diagonal matrix consisting of the diagonal entries of $ \mat A $.
Written in this way, $\mat{B}$ is a random matrix with mean zero and null diagonal.
We will show that the bulk of the spectrum of $\mat M$ is equivalent to the bulk of eigenvalues of the matrix $ \mat{J} =  \mat{X} (\mat{D} - \mu \mat{I}+ \mat{B}) $.

Using equation~\ref{eq:matrix_decomposed}, the resolvent of $\mat M$ can be written 
\begin{equation} 
\G(\quat{q}; \mat{M}) = \mathbb{E} [\frac{1} {S} \tr \left( \quat{q} \mat{I} - \mat{X} \mat{D} - \mu \mat{X} - \mat{X} \mat{B} - \mu \mat{X} \mat{1} \right)^{-1}] 
= \mathbb{E} [\frac{1} {S} \tr \left( \quat{q} \mat{I} - \mat{J} - \mu \mat{X} \mat{1} \right)^{-1}].
 \end{equation}

\begin{figure}
	\includegraphics[scale = .75] {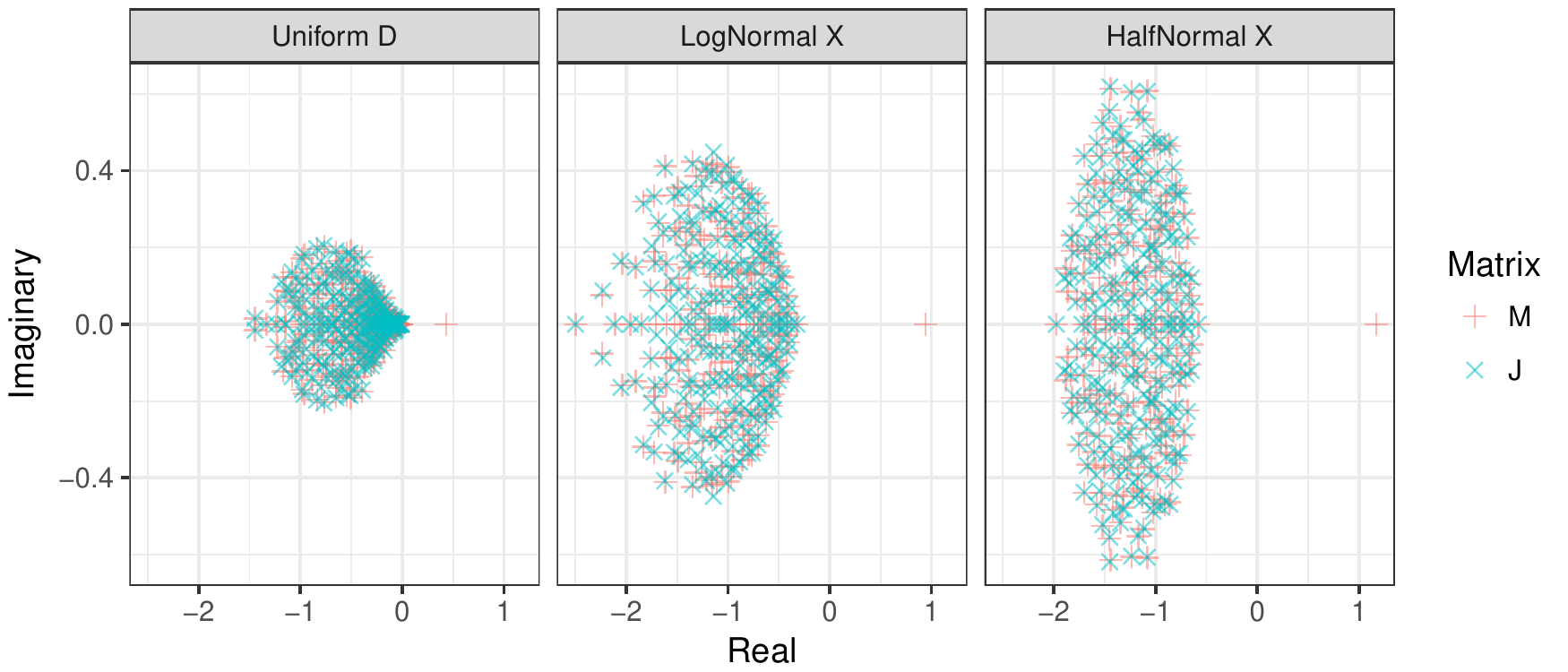}
	\caption{The eigenvalue distributions for $\mat M $ and $ \mat J $ of size $ S = 250 $ with $\mat D $ and $\mat X$  following different distributions. In each case, $\mat A $ is a bivariate normal distribution with identical marginals $ \mu = 2 / S$, $\sigma = 0.5$ and $ \rho = 0$. Uniform D has $\mat D $ following a uniform distribution on $(-1.5, -0.5)$. LogNormal X has $\mat X$ following a log-normal distribution with log-mean $0$ and log-standard deviation $0.35$. HalfNormal X has $ \mat X $ following a half-normal distribution with support $(1, \infty)$ and parameter $\theta = 1$.}
	\label{fig:M+J}
\end{figure}

Using the Sherman-Morrison formula, if $\mat{Y}$ and $\mat{Y + Z}$ are invertible matrices and $\mat{Z}$ has rank 1, then
\begin{equation}
(\mat{Y} + \mat{Z})^{-1} = \mat{Y}^{-1} + \frac{1}{1+ \tr (\mat{Z}\mat{Y}^{-1})} \mat{Y}^{-1} \mat{Z} \mat{Y}^{-1} \ . 
\end{equation}
Since $\mu \mat{X} \mat{1}$ has rank one, we have
\begin{equation}
 \left( \quat{q} \mat{I} - \mat{J} - \mu \mat{X}\mat{1} \right)^{-1} = (\textbf{q} \mat{I} - \mat{J})^{-1} +
\frac{1} {1+ \tr (\mu \mat{X}\mat{1} (\textbf{q} \mat{I} - \mat{J})^{-1}) } (\textbf{q} \mat{I} - \mat{J})^{-1} \mu \mat{X}\mat{1} (\textbf{q} \mat{I} - \mat{J})^{-1} \ .
 \label{inverse}
 \end{equation}
By introducing the linear operator $\langle \cdot \rangle$ defined by $\langle \mat {C} \rangle = \frac{1} {S} \tr {\mat{C}} $ for an $S \times S $ matrix $\mat C$,
we obtain
 \begin{equation}
 \G(\quat{q}; \mat{M})  = \G(\quat{q}; \mat{J}) +  \frac{\mu}{1+ S \mu \langle \mat{X}\mat{1} (\textbf{q} \mat{I} - \mat{J})^{-1}) \rangle }
 \langle (\textbf{q} \mat{I} - \mat{J})^{-1} \mat{X} \mat{1} (\textbf{q} \mat{I} - \mat{J})^{-1} \rangle  \ .
  \label{eq:subleading}
\end{equation}
In the limit of large $S$, the contribution from the second term in~\ref{eq:subleading} is subleading. Therefore, the resolvent of $\mat M$ converges to the resolvent of $\mat J$ when $S$ is large.
In other words,  as shown in Figure \ref{fig:M+J}, the bulks of the eigenvalues of $\mat{M}$ and $\mat{J}$ are the same --- up to finite-size corrections.


\section{The case $\sigma = 0$}

When $\sigma = 0$, we derive the spectrum of a matrix $\mat{Q} = \mat{X} \left(  \mat{D} + \mu \mat{1} \right)$. This case corresponds to
setting $\mat{B} = 0$ in equation~\ref{eq:matrix_decomposed}.
As shown in the main text, the spectral density of the matrix $\mat Q $ is characterized by the presence of an outlier. In this section, we focus on the
bulk of eigenvalues.

\begin{figure}
	\includegraphics[scale = .75] {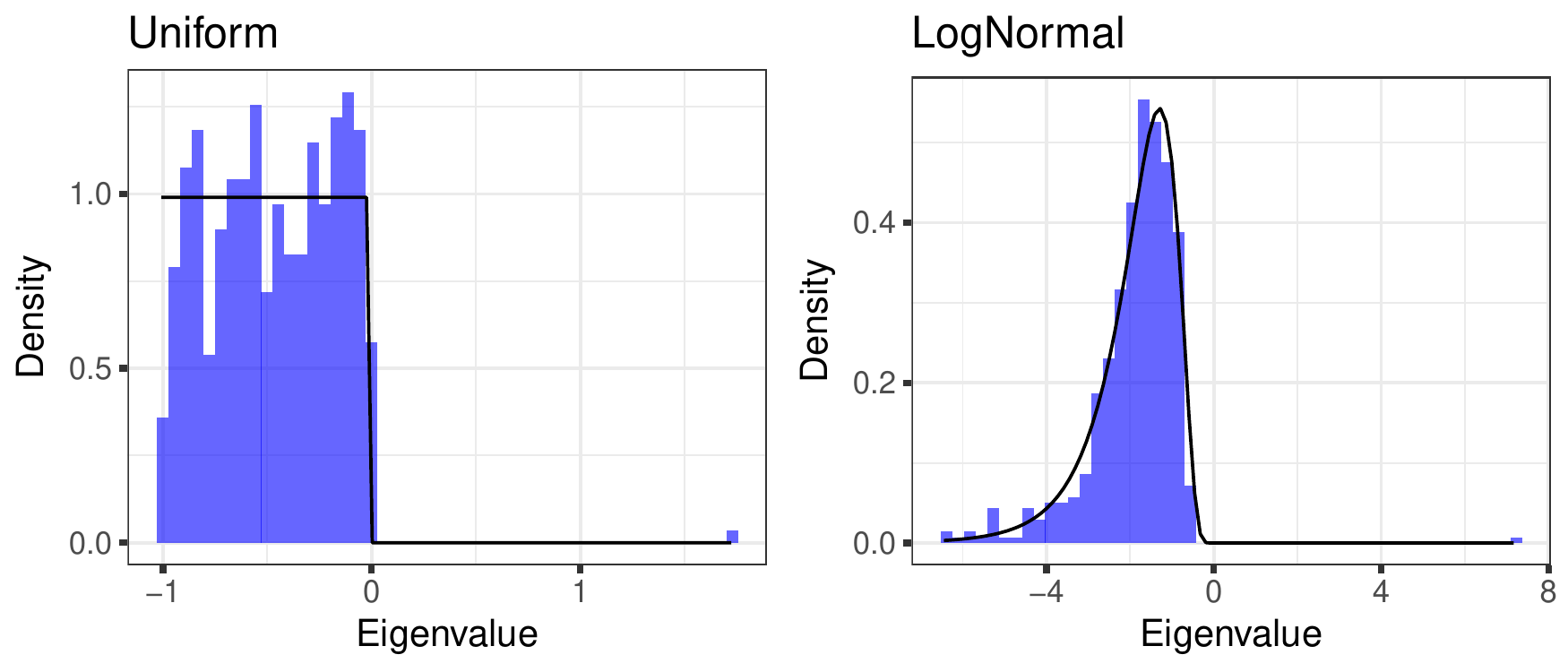}
	\caption{Histograms of the eigenvalue distribution for two matrices $\mat Q$ of size $ S = 1500$. In each matrix, $\sigma_d = 0$, $\mu_d = -1$ and $ \mu = 5 / S$. The Uniform plot has $\mat{X}$ following a uniform distribution on $(0,1)$ and the LogNormal plot has $\mat{X}$ following a log-normal distribution with log-mean and log-standard deviation both $0.5$.}
	\label{fig:observationSI}
\end{figure}

If we take $ \mat{J} =  \mat{X} ( \mat{D} - \mu \mat{I} + \mat{B}) $ as before and set $\sigma = 0$, then $\mat B = 0 $, so that $\mat {M} = \mat{Q}$ and $ \mat{J} =  \mat{X} ( \mat{D} - \mu \mat{I})$.  The bulk of eigenvalues of $\mat{Q}$ and  $\mat{J}$ will be the same. The resolvent of $\mat{J}$ in the case $\sigma = 0$ reads
\begin{equation}
\begin{aligned}
G(\textbf{q}; \mat{J}) & = \frac{1} {S} \tr \left( \textbf{q} - \mat{J} \right)^{-1} \\ 
& = \frac{1} {S} \tr \left( \textbf{q} - \mat{X}  \mat{D}   \right)^{-1} \\
& = \frac{1} {S} \sum_{i = 1}^S \frac{1} {q - X_i D_i } \\
\label{resolventQ}
\end{aligned}
\end{equation}
since $\textbf{q} - \mat{X} ( \mat{D} - \mu \mat{I})$ is symmetric. In the limit of large $S$, the sum in eq.~\ref{resolventQ} tends toward $ \mathbb{E} \left[(\textbf{q} - \mat{X} \mat{D} )^{-1} \right]$. If $P_{XD}(x,s)$ is the joint distribution of the entries of $\mat{X}$ and $\mat{D}$, we obtain
\begin{equation}
\begin{aligned}
G(\textbf{q}; \mat{J}) & = \frac{1} {S} \sum_{i = 1}^S \frac{1} {q - X_iD_i + \mu X_i} \\
& = \mathbb{E} \left[(\textbf{q} - \mat{X} ( \mat{D} - \mu \mat{I}))^{-1} \right] \\
& = \int dx \ ds \ \frac{P_{XD}(x,s)} {q - x s} \ . 
\label{spectrumQ}
\end{aligned}
\end{equation}
In the case of a constant diagonal matrix $\mat{D} = d \mat{I}$, this equation simplifies to
\begin{equation}
G(\textbf{q}; \mat{J}) = \int d x \ \frac{P_X(x)} {q - x d } 
= \frac{1}{d}\int d y \ \frac{P_X\left(\frac{y}{d} \right)}{q - y}  \\
\end{equation}
with the change of variables $y = x d$. Using equation~\ref{resolvent+spectrum}, we obtain that the spectral density $\varrho_{\mat{J}}(\lambda)$ will be 
\begin{equation}
\varrho_{\mat{J}}(\lambda) =  \frac{1}{d} P_X \left(\frac{\lambda} { d} \right) \ .
\label{predictedQ}
\end{equation}
In Figure \ref{fig:observationSI}, we plot the prediction from Equation \ref{predictedQ} against the bulk of the spectrum of $\mat Q $ for two distributions of $\mat X$.

\section{Derivation of the spectral density using the cavity method}

In section~\ref{sec:mudiff}, we showed that we can isolate the effect of $\mu \neq 0$. In this section, we use the cavity method~~\cite{Rogers2008,Rogers2009} to
derive the
spectrum of the matrix $ \mat{J} =  \mat{X} ( \mat{D} - \mu \mat{I} + \mat{B}) $, where $\mat{D}$ and $\mat{X}$ are two random diagonal
matrices and $\mat{B}$ is a random matrix following the elliptic law.

We introduce the resolvent matrix 
\begin{equation}
\mat{G} = ( \mat{M} - \quat{q} \mat{I}  )^{-1} \ ,
\end{equation}
The resolvent can be written as
\begin{equation}
\G(\quat{q}; \mat{B}) = \frac{1} {S} \tr \mat{G}  \ .
\end{equation}
Note that each element of the resolvent matrix is a quaternion. In particular we will use the notation
\begin{equation}
\quat{G}_{ik} = \alpha_{ik} + \beta_{ik} j \equiv \left( \begin{matrix} \alpha_{ik} & \beta_{ik} \\ \bar{\beta_{ik}} & \bar{\alpha_{ik}} \\ \end{matrix} \right) \ ,
\end{equation}
while $\G = \alpha + \beta j$, where $\alpha = \sum_i \alpha_{ii} / S$ and $\beta = \sum_i \beta_{ii} / S$.
The cavity method allows us to compute the elements of $\mat{G}$ (and therefore the resolvent $\G$) if the matrix $\mat{M}$ has a tree structure~\cite{Rogers2008,Rogers2009}.
It also allows to compute the spectral density for large, densely connected, random matrices~\cite{Rogers2008,Rogers2009,Grilli2016}.
In the limit of large $S$, for a densely connected matrix $\mat{M}$, the cavity equations read~\cite{Rogers2009,Grilli2016}
\begin{equation}
\quat{G}_{il} \equiv 
\left( \begin{matrix} \alpha_{il} & \beta_{il} \\ \bar{\beta}_{il} & \bar{\alpha}_{il} \\ \end{matrix} \right) = 
- \left(
\left( \begin{matrix} \lambda & \epsilon \\ \epsilon & \bar{\lambda} \\ \end{matrix} \right) 
+ \sum_{jk}
\left( \begin{matrix} M_{ij} & 0 \\ 0 & M_{ji} \\ \end{matrix} \right)
\left( \begin{matrix} \alpha_{jk} & \beta_{jk} \\ \bar{\beta}_{jk} & \bar{\alpha}_{jk} \\ \end{matrix} \right)
\left( \begin{matrix} M_{kl} & 0 \\ 0 & M_{lk} \\ \end{matrix} \right)
\right)^{-1}
 \ .
\end{equation}
By introducing 
\begin{equation}
\matquat{M}_{ij} = \left( \begin{matrix} M_{ij} & 0 \\ 0 & M_{ji} \\ \end{matrix} \right)
 \ ,
\end{equation}
we obtain the more compact equation
\begin{equation}
\quat{G}_{il} = -\left(
\quat{q} + \sum_{jk} \matquat{M}_{ij} \quat{G}_{jk} \matquat{M}_{kl} \right)^{-1} \ .
\end{equation}

Our goal is to find the resolvent for a random matrix of the form $\mat{M} = \mat{X} (\mat{D} + \mat{B})$,
where $\mat{X}$ and $\mat{D}$ are diagonal matrices, while $\mat{B}$ is a random matrix following the elliptic law.
We introduce the matrix
\begin{equation}
 \left( \begin{matrix}
-\quat{q} \mat{I} & \matquat{X} \\
- \matquat{D} - \matquat{B} & \mat{I}
\end{matrix} \right) \ ,
\end{equation}
which, when quaternions are represented as $2 \times 2$ matrices, is a 
$4S \times 4S$ matrix. In particular,  this matrix is composed of $S^2$
$4 \times 4$ blocks with entries
\begin{equation}
 \left( \begin{matrix}
-\lambda \delta_{ij} & -\epsilon \delta_{ij} & X_{ii} \delta_{ij} & 0 \\
-\epsilon \delta_{ij} & -\bar{\lambda} \delta_{ij}  & 0 & X_{ii} \delta_{ij} \\
-D_{ii} \delta_{ij} - B_{ij} & 0 &  \delta_{ij} & 0 \\
0 & -D_{ii} \delta_{ij} - B_{ji} & 0 &  \delta_{ij}  
\end{matrix} \right) \ .
\end{equation}

It is simple to observe that
\begin{equation}
 \mat{T} = \left( \begin{matrix}
-\bf{q} & \mat{X} \\
\mat{- A} & \mat{I}
\end{matrix} \right)^{-1} 
= \left( \begin{matrix} 
\left( -\quat{q} \mat{I} + \mat{X} (\mat{D} + \mat{B}) \right)^{-1}  & ... \\
... & ...
\end{matrix} \right)
= \left( \begin{matrix} 
\mat{G} & ... \\
... & ...
\end{matrix} \right) \ .
\end{equation}

If we write the cavity equation for $\mat{T}$, assuming dense matrices, we obtain
\begin{equation}
\quat{T}_{ii} = \left[ \left( \begin{matrix}
-\mat{q}  & \matquat{X}_{ii} \\
- \matquat{D}_{ii} & 1
\end{matrix} \right)
- \sum_{j,k }
\left( \begin{matrix}
\mat{0} & \mat{0} \\
\matquat{B}_{ij} & \mat{0}
\end{matrix} \right)
\mat{T}_{jk}
\left( \begin{matrix}
\mat{0} & \mat{0} \\
\matquat{B}_{ki} & \mat{0}
\end{matrix} \right)
\right]^{-1} \ .
\end{equation}

 We can apply the law of large numbers to take the expectation of the matrices over $\mat{B}$.
Using the following expectations over the elements of $\mat{B}$
\begin{equation}
\begin{split}
\mathbb{E}\left[ B_{ij} \right] &= 0 \\
\mathbb{E}\left[ (B_{ij} )^2 \right] & = \frac{\tilde{\sigma}^2}{S} \\
\mathbb{E}\left[ B_{ij} B_{ji} \right] & = \rho \frac{\tilde{\sigma}^2}{S} \ ,
\end{split}
\end{equation}
we find that the non-diagonal terms of $\mat{T}$ go to zero in expectation.
By introducing the notation
\begin{equation}
\quat{T}_{ii}  \equiv
\left(
	\begin{array} {cc}
	\quat{G}_{ii} & \quat{T}^b_{ii} \\
	... & ...
	\end{array}
	\right) 
\end{equation}
and
\begin{equation}
\quat{T}_{\star} \equiv \frac{1}{S} \sum_i \quat{T}_{ii} =
\left(
	\begin{array} {cc}
	\G & \quat{T}^b_{\star} \\
	... & ...
	\end{array}
	\right) \ ,
\end{equation}
the cavity equations for the diagonal terms read
\begin{equation}
\quat{T}_{ii} 
= \left(
	\begin{array} {cc}
	-\quat{q} & X_{ii} \\
	- D_{ii} - \quat{t} \circ \quat{T}^b_\star & \mat{I}
	\end{array}
	\right)^{-1}
\end{equation}
where $\quat{T}_\star = \sum_i \quat{T}_{ii} / S $,
while $\bf{t} = \tilde{\sigma}^2\rho + \tilde{\sigma}^2j$ with $\circ$ denoting the element-by-element matrix product (see section~\ref{sec:quaternions}).

We obtain a system of $2S$ quaternionic equations
\begin{equation}
\begin{cases}
\displaystyle
\quat{T}^b_{ii} = - X_{ii} \left(-\quat{q} -X_{ii} \left( - D_{ii} - \quat{t} \circ \quat{T}^b_\star \right) \right)^{-1} \\
\displaystyle
\quat{G}_{ii} =   \left( - \quat{q} - X_{ii} \left(- D_{ii} - \quat{t} \circ \quat{T}^b_\star  \right) \right)^{-1} \ ,
\end{cases}
\end{equation}
where 
\begin{equation}
\quat{T}^b_\star = \frac{1}{S} \sum_i \quat{T}_{ii} \ .
\end{equation}

Assuming that the elements of $\mat{X}$ and $\mat{D}$ are drawn from a given joint distirbution $P_{XD}$, we obtain the following 
two equations
\begin{equation}
\begin{cases}
\displaystyle
\quat{T}^b_{\star} = - \int dx \ ds \ P_{XD}(x,s)  x \left(-\quat{q} + x (  s + \quat{t} \circ \quat{T}^b_\star ) \right)^{-1} \\
\displaystyle
\G = \int dx \ ds \ P_{XD}(x,s)  \left( - \quat{q} + x ( s + \quat{t} \circ \quat{T}^b_\star ) \right)^{-1} \ .
\end{cases}
\label{eq:tbstar}
\end{equation}

At this point one can use
 $\quat{T}^b_{\star}= \alpha^b + \beta^b j $ and $\G = \alpha + \beta j$
to obtain $4$ equations of complex variables. It is simple to observe that $\beta^b = 0$ is always a solution and
$\beta = 0$ if and only if $\beta^b = 0$. The solution $\beta = \beta^b = 0$ always corresponds to a null specral density~\cite{Rogers2010}.
The values of $\lambda$
for which a non-zero solution for $\beta$ exist correspond to the support of the spectral density.
Setting $\beta^b \neq 0$ and $\epsilon = 0$, one obtains from equation~\ref{eq:tbstar}
\begin{equation}
\begin{cases}
\displaystyle
\alpha^b =  \int dx \ ds \ P_{XD}(x,s)  
 \frac{ x \left( \bar{\lambda}  - s x - x \tilde{\sigma}^2 \rho \bar{\alpha}^b \right) }{
  | \lambda + x \left( -s-\alpha^b \rho\tilde{\sigma}^2 \right) | ^2 + | x\tilde{\sigma}^2 \beta^b |^2 } \\
\displaystyle
1 = \int dx \ ds \ P_{XD}(x,s)  \frac{ x^2 \tilde{\sigma}^2} {| \lambda + x \left( -s-\alpha^b \rho\tilde{\sigma}^2 \right) | ^2 + | x\tilde{\sigma}^2 \beta^b |^2 } \ .
\end{cases}\label{support:2}
\end{equation}
By using the second equation, the system of equations further simplifies to
\begin{equation}
\begin{cases}
\displaystyle
\alpha^b + \rho \bar{\alpha}^b =  \int dx \ ds \ P_{XD}(x,s)  
 \frac{ x \left( \bar{\lambda}  - s x  \right) }{
  | \lambda + x \left( -s-\alpha^b \rho\tilde{\sigma}^2 \right) | ^2 + | x\tilde{\sigma}^2 \beta^b |^2 } \\
\displaystyle
1 = \int dx \ ds \ P_{XD}(x,s)  \frac{ x^2 \tilde{\sigma}^2} {| \lambda + x \left( -s-\alpha^b \rho\tilde{\sigma}^2 \right) | ^2 + | x\tilde{\sigma}^2 \beta^b |^2 } \ .
\end{cases}\label{support:3}
\end{equation}

The values of $\lambda$ for which a solution of equations~\ref{support:3} exists are contained in the support
of the spectral density.
We assume that the solution $\beta^b$ of these equations vanishes at the boundaries of the support. In this case, the
points at the boundaries of the support of the spectral density are the complex solutions $\lambda$ of
\begin{equation}
\begin{cases}
\displaystyle
\alpha^b +  \rho \bar{\alpha}^b  =  \int dx \ ds \ P_{XD}(x,s)  
 \frac{ x \left( \bar{\lambda}  - s x \right) }{
  | \lambda + x \left( -s-\alpha^b \rho\tilde{\sigma}^2 \right) | ^2  } \\
\displaystyle
1 = \int dx \ ds \ P_{XD}(x,s)  \frac{ x^2 \tilde{\sigma}^2} {| \lambda + x \left( -s-\alpha^b \rho\tilde{\sigma}^2 \right) | ^2 } \ ,
\end{cases}
\label{eq:support}
\end{equation}
where we used the second equation

\noindent We should note here that this method does not prove the
convergence in any mode of the spectral bulk,
but does yield a prediction for its support.

\section*{Support of the spectral density in the case $\rho = 0$}

In the case $\rho = 0$, the two equations~\ref{eq:support} become independent, and the support is defined by the solutions of
\begin{equation}
1 = \tilde{\sigma}^2 \int dx \ ds \ P_{XD}(x,s)  \frac{ x^2 } {| \lambda - s x  | ^2 } \ .
\end{equation} 
In the case $\tilde{\sigma}_d^2 = 0$, this equation further simplifies to
\begin{equation}
1 = \tilde{\sigma}^2 \int dx \ P_{X}(x)  \frac{ x^2 } {| \lambda - \mu_d x  | ^2 } \ . 
\end{equation} 
For instance, if $P_{X}$ is a uniform distribution on $[0,1]$, one can evaluate the integral, obtaining
\begin{equation}
\begin{aligned}
 1 & = \int_0^1 dx \ \frac{ x^2 \tilde{\sigma}^2} {| \lambda - \mu_d x | ^2 } \\
   &= \frac{ \tilde{\sigma} ^ 2} { \mu_d^3} \left( 2 \lambda \mu_d - \mu_d ^ 2 + 2 \lambda (\mu_d - \lambda) \log \left| \frac{\lambda} {\lambda - \mu_d} \right| \right) \ .
 \end{aligned}
 \end{equation}
The complex solutions $\lambda$ of this equation define the boundary of the support of the spectral density.

\begin{figure} [p]
	\includegraphics[scale = .75] {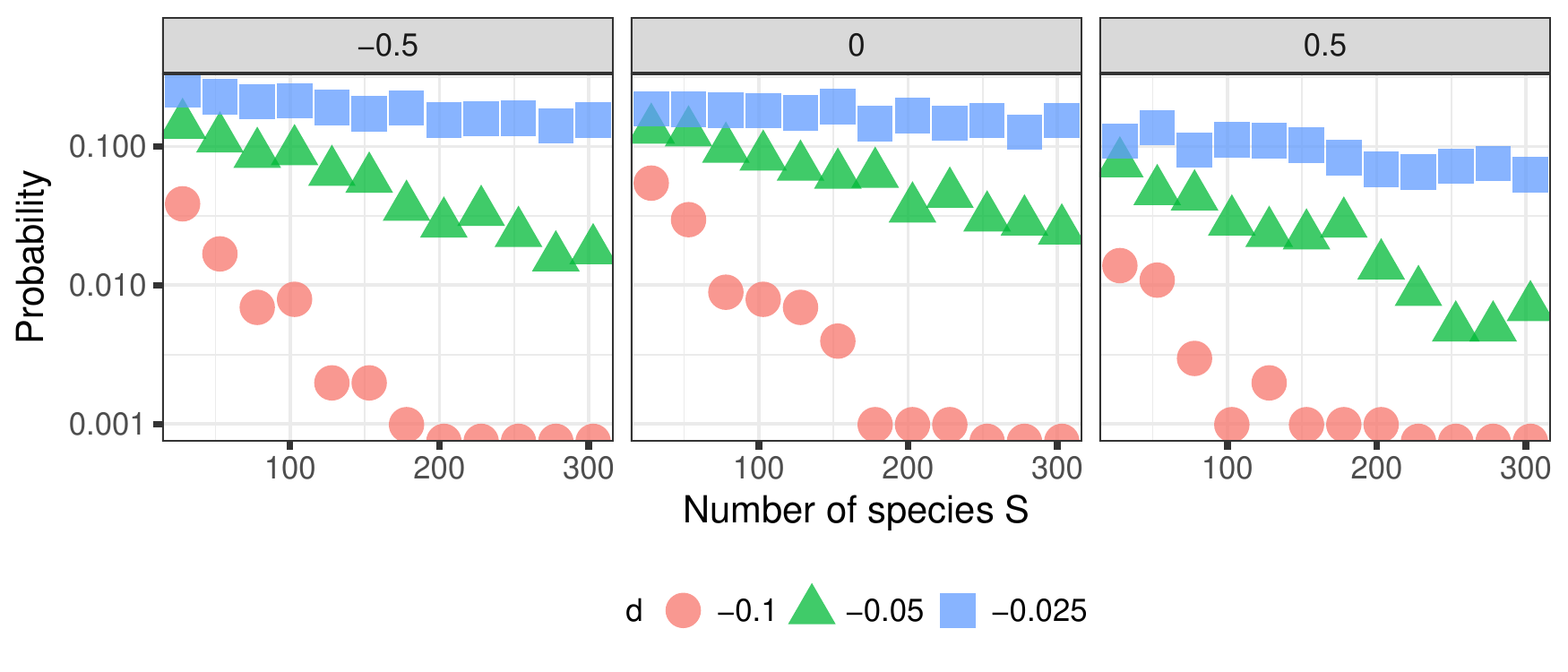}
	\caption{The conditional probability of a matrix $\mat M$ becoming \emph{stable} given an \emph{unstable} interaction matrix $\mat A$ with leading eigenvalue $-d$. The off-diagonal elements of $\mat A$ follow a bivariate normal distribution and, for each $S$, $\mu$ =0, $\sigma = 1 / \sqrt{S}$ and $\rho = -0.5$, $0$ or $0.5$. The elements of $\mat X$ are drawn from a uniform distribution on $(0,1)$, so that $\mu_X = 0.5$ and $\sigma_X^2 = \frac{1}{12}$. The diagonal elements of $\mat A$ are fixed at $-1$ since $\sigma_d^2 = 0$ and $\mu_D = -1$. Each probability is calculated from $1000$ trials from $1000$ simulated matrices $\mat{M} = \mat{XA}$.}
	\label{fig:destabilized1}
\end{figure}

\begin{figure}
	\includegraphics[scale = .75] {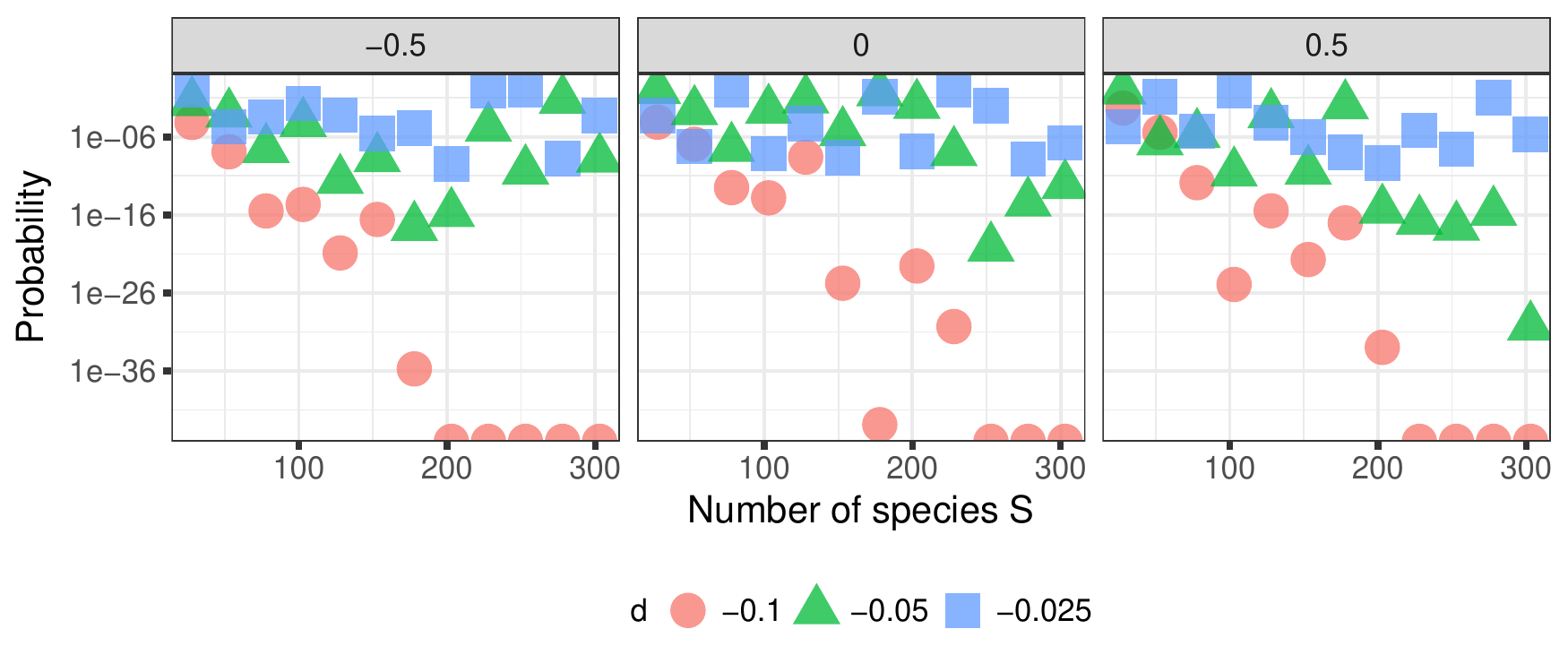}
	\caption{As in Figure \ref{fig:destabilized1}, except now the conditional probability is weighted by feasibility.}
	\label{fig:destabilized2}
\end{figure}

\begin{figure}
	\includegraphics[scale = .75] {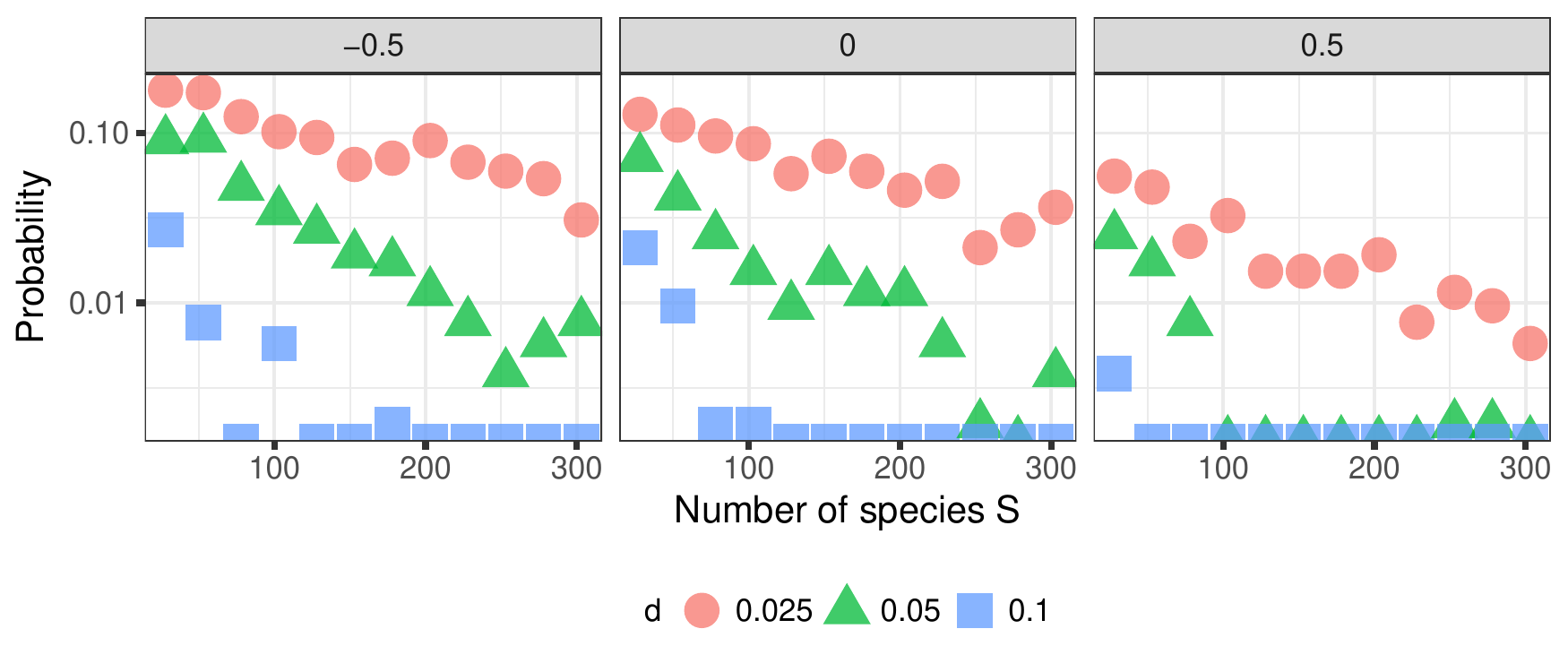}
	\caption{The conditional probability of a matrix $\mat M$ becoming unstable given a stable interaction matrix $\mat A$ with leading eigenvalue $-d$. The off-diagonal elements of $\mat A$ follow a bivariate normal distribution and, for each $S$, $\mu$ =0, $\sigma = 1 / \sqrt{S}$ and $\rho = -0.5$, $0$ or $0.5$. The elements of $\mat X$ are drawn from a uniform distribution on $(0,1)$, so that $\mu_X = 0.5$ and $\sigma_X^2 = \frac{1}{12}$. The diagonal elements of $\mat A$ follow a uniform distribution on $(-0.75, -1.25)$ so that $\sigma_d^2 = \frac{1}{48}$ and $\mu_D = -1$. Each probability is calculated from $1000$ trials from $1000$ simulated matrices $\mat{M} = \mat{XA}$.}
	\label{fig:Dvarying1}
\end{figure}

\begin{figure}
	\includegraphics[scale = .75] {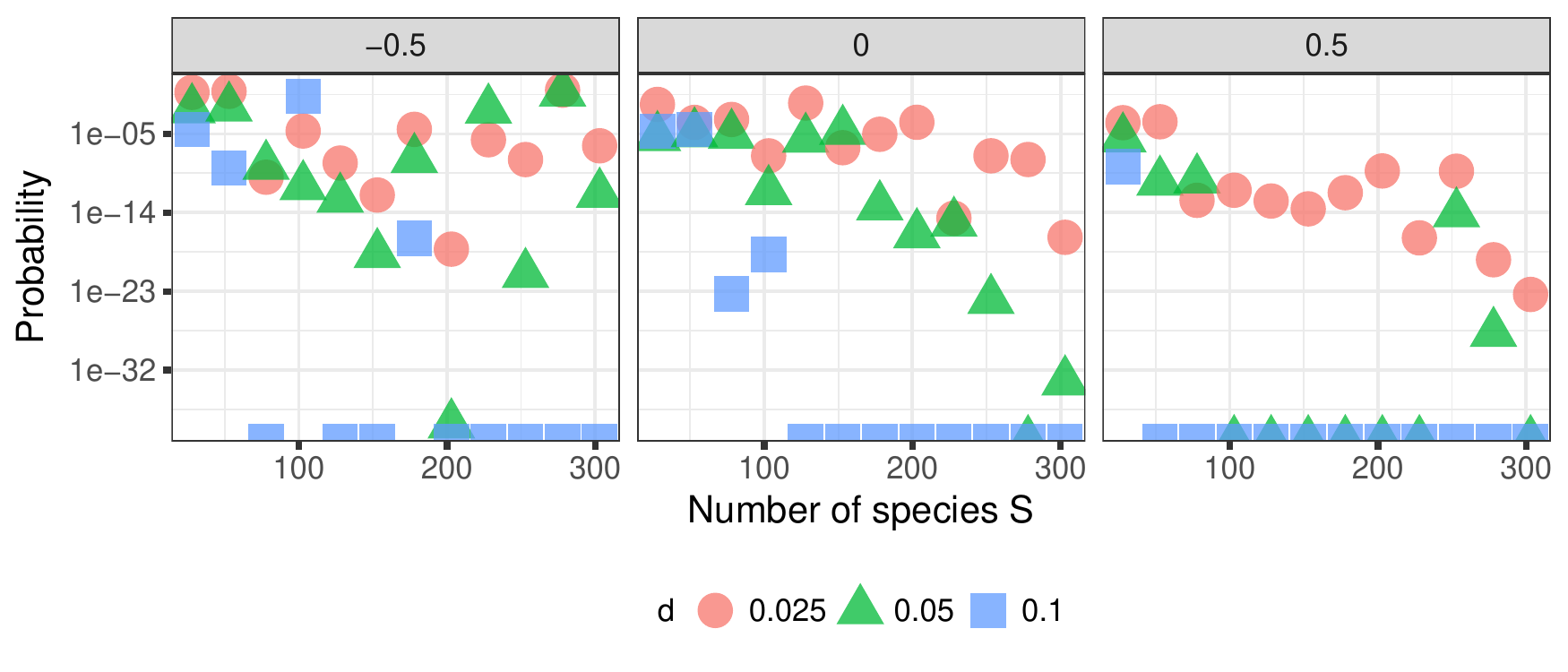}
	\caption{As in Figure \ref{fig:Dvarying1}, except now the conditional probability is weighted by feasibility.}
	\label{fig:Dvarying2}
\end{figure}

\end{document}